\documentclass[final,5p,times]{elsarticle}

\usepackage{lineno}
\usepackage{amsmath,amssymb,amsfonts}
\usepackage{graphicx}
\usepackage{textcomp}
\usepackage{xcolor}
\usepackage{booktabs}
\usepackage{multirow}
\usepackage{microtype}
\usepackage{xurl}      
\usepackage[hidelinks]{hyperref}

\journal{Neurocomputing}
\renewcommand{\arraystretch}{1.22}
\usepackage{float}        
\usepackage{dblfloatfix}   
\usepackage{placeins}      

\renewcommand{\arraystretch}{1.22}
\usepackage{float}        
\usepackage{dblfloatfix}   
\usepackage{placeins}      

\begin{document}

\begin{frontmatter}

\title{The Same Ledger, Different Verdicts: How Measurement
Specification Determines On-Chain Concentration}

\author[a]{Jintao Liu\corref{cor}\fnref{eq}}
\ead{LIU.10@student.unimelb.edu.au}

\author[a]{Zhimo Ji\fnref{eq}}
\ead{zhimoji@student.unimelb.edu.au}

\author[b]{Xuzhe Lin}
\ead{xlin0415@uni.sydney.edu.au}

\cortext[cor]{Corresponding author.}
\fntext[eq]{These authors contributed equally to this work.}

\affiliation[a]{organization={School of Computing and Information Systems,
                              The University of Melbourne},
                city={Melbourne},
                country={Australia}}

\affiliation[b]{organization={School of Mathematics and Statistics,
                              The University of Sydney},
                city={Sydney},
                country={Australia}}

\begin{abstract}
Whether a public blockchain is ``decentralized'' is routinely settled by
citing a concentration statistic. We show that on two ERC-20 ledgers---the
Chainlink (LINK) oracle token and the Uniswap (UNI) governance token,
observed over an identical 90-day window---that verdict is not a property of
the ledger but of the measurement. Four discretionary choices (holder
population, address type, temporal aggregation, and entity resolution) move
the balance Herfindahl--Hirschman Index for UNI from 109 to 2,336, a factor
of 21, with every specification defensible and none dictated by the data.
Over the same range the Gini coefficient moves by less than $0.003$, and
under entity resolution not at all to four decimal places, so the two
statistics are not substitutes and the choice between them is itself
consequential. We further document a case in which a single
implementation detail---whether repeated transfers between an address pair
are summed or overwritten---overturns a finding about wealth and
structural position that had been written into an earlier draft of this
paper. Substantively, both
ledgers are extraordinarily unequal in ownership (balance $Gini = 0.990$ and
$0.998$) yet unconcentrated in routing (weekly $HHI_{\text{Flow}} = 421$ and
$386$ at address level, $654$ and $667$ after partial entity resolution), and
across addresses the two dimensions are close to statistically independent.
Applying a parameterised criterion for \emph{hidden brokers}---structurally
central, zero-balance, carrying no public label---we recover 30 and 18 such
addresses, twelve shared across both ledgers and half of those resolving to no
named protocol under manual on-chain verification; a matched control
experiment shows the criterion, not the representation used to operationalise
it, does the work. Finally, on the governance ledger the addresses that can
amend protocol rules and those that route the token are almost disjoint, so
whatever contestability the routing layer exhibits is held at the pleasure of
a rule layer with a Nakamoto coefficient of two. We conclude that on-chain
concentration should be reported as a specified range rather than a point
estimate, and give the specification set required for comparability.
\end{abstract}

\begin{highlights}
\item Four discretionary measurement choices---three governing the holder
population, one the temporal horizon---move the balance HHI of a single
ERC-20 ledger from 109 to 2,336, a factor of 21, with no specification
dictated by the data.
\item The Gini coefficient moves by less than $0.003$ over that same range and,
under entity resolution, not at all to four decimal places; Gini and HHI
answer different questions and cannot substitute for one another.
\item One implementation detail---summing versus overwriting repeated
transfers between an address pair---overturns a finding about wealth and structural position that had
been written into an earlier draft of this paper.
\item Ownership is extraordinarily unequal on both ledgers ($Gini = 0.990$ and
$0.998$) while routing is unconcentrated ($HHI_{\text{Flow}} = 421$ and
$386$); across addresses the two dimensions are close to independent.
\item Routing contestability is revocable: on the governance ledger the
addresses that can amend the rules and those that route the token are almost
disjoint, and the rule layer has a Nakamoto coefficient of two.
\end{highlights}


\end{frontmatter}


\section{Introduction}
\label{sec:intro}

The advent of public blockchain networks---most prominently Ethereum---has given monetary economists and quantitative sociologists an unusually rich empirical setting: a ``public laboratory'' in which every transaction, asset transfer, and contract interaction across an entire monetary ecosystem is immutably recorded on a transparent, permissionless ledger. Early whitepapers \cite{nakamoto2008bitcoin, wood2014ethereum} envisioned these ledgers as engines of broad disintermediation and financial democratization, capable of eliminating institutional gatekeepers and replacing centralized trust with cryptographic consensus (``Code is Law'' \cite{lessig1999code}).

As decentralized finance (DeFi) has matured into a multi-billion-dollar global clearing infrastructure, however, empirical observations increasingly reveal a tension that deserves scrutiny: the emergence of systemic re-centralization. Far from yielding a flat, egalitarian economic topology, on-chain ledger distributions routinely display extreme wealth polarization and structural bottlenecks.

Uncovering the micro-structural mechanics of this re-centralization presents a non-trivial computational problem. Traditional blockchain analysis relies heavily on static wealth heuristics or classical graph centrality metrics (e.g., degree, betweenness, and PageRank) \cite{meiklejohn2013fistful}. In massive, sparse, and dynamic smart contract interaction graphs, these metrics face computational bottlenecks---scaling at $O(|V| \cdot |E|)$ \cite{brandes2001faster}---and fail to capture subtle structural roles. \textbf{Graph Neural Networks (GNNs)} offer a more expressive alternative \cite{kipf2017semi, velickovic2018graph, hamilton2017inductive}: by aggregating higher-order topological neighborhoods and localized transactional features into low-dimensional node embeddings, they enable the unsupervised discovery of functional economic roles and zero-balance structural intermediaries that would otherwise remain buried in raw ledger data \cite{weber2019anti}.

A separate but equally important concern is the choice of benchmark. Previous empirical studies evaluating blockchain inequality rely predominantly on global wealth distribution metrics---such as the Gini coefficient or Top-$k$ shares---frequently benchmarking token ledgers against sovereign nation-states \cite{gudgeon2020defi, campajola2023evolution, zarir2021developing}. This creates a conceptual mismatch: while a sovereign nation is a multi-sectoral civil society encompassing public infrastructure, real estate, and labor markets, a smart contract token ledger functions primarily as specialized financial infrastructure---a price-feed and clearing layer in the case of an oracle token, a governance and routing layer in the case of an exchange token. Evaluating dedicated financial rails using macro-social wealth metrics risks mischaracterizing market-driven liquidity aggregation as a failed social experiment.

To bridge graph neural representation learning with monetary economic theory, this paper presents a unified empirical and GNN-driven framework applied to \emph{two} functionally dissimilar ERC-20 ledgers over an identical 90-day window: the Chainlink (LINK) oracle token ($642{,}126$ transfer-active addresses) and the Uniswap (UNI) governance token ($38{,}667$). A single-asset study cannot separate a property of one protocol from a property of permissionless ledgers in general; holding the observation period fixed removes market-regime effects as a confound, so any divergence between the two is attributable to the asset rather than to the period. The comparison is adversarial to our own thesis, since UNI's entire design premise is broadly distributed community control.

Combining unsupervised GNN node embeddings with concentration metrics computed at two scales---ownership (Gini, top-$k$, HHI) and routing (weekly flow HHI)---we decode the divergence between static wealth and dynamic structural power. We evaluate five research questions:
\begin{itemize}
    \item \textbf{RQ1 (Macro vs.\ industrial concentration)}: \textit{How severe is on-chain inequality when benchmarked concurrently against sovereign wealth distributions (Gini) and industrial market concentration thresholds (HHI), and does the answer replicate across functionally dissimilar tokens?}
    \item \textbf{RQ2 (Structural power vs.\ wealth)}: \textit{Do learned graph representations reveal ``hidden brokers'' that occupy high structural influence despite holding no token balance, and are such brokers specific to a token or shared across ledgers?}
    \item \textbf{RQ3 (Role typology and functional homology)}: \textit{What economic roles emerge from unsupervised node embeddings, and how does the resulting clearing topology compare to traditional financial intermediaries?}
    \item \textbf{RQ4 (Topological resilience)}: \textit{How does structural concentration respond to an extreme exogenous shock, and does it recover?}
    \item \textbf{RQ5 (Measurement robustness and the rule layer)}: \textit{How sensitive are these concentration verdicts to the discretionary choices a measurement requires---population, aggregation, and above all entity resolution---and does a token-weighted governance layer retain the authority to revoke whatever contestability the routing layer exhibits?}
\end{itemize}
Section~\ref{sec:rq} returns to each of these in light of the evidence.

\subsection{Summary of Main Contributions}
This paper makes three contributions to the measurement of on-chain
concentration.

\subsubsection{Specification Dependence of On-Chain Concentration (C1)}
\label{sec:contrib-spec}
Our primary contribution is to show that the headline verdict---whether a
ledger is ``concentrated''---is determined less by the ledger than by four
discretionary choices the analyst must make and that no data can settle:
which holders count, which address types are excluded, over what horizon
flows are aggregated, and whether an address or an entity is the unit of
analysis. On UNI these choices move the balance Herfindahl--Hirschman Index
from $109$ to $2{,}336$, a factor of $21$; on LINK, from $115$ to $562$.
Every one of these specifications is defensible and several appear in the
published literature. Over the same range the Gini coefficient moves by less
than $0.003$ and, under entity resolution, not at all to four decimal
places---so the two statistics are not substitutes, and a study reporting
one of them has not thereby also reported the other.

\begin{figure}[H]
  \centering
  \includegraphics[width=\linewidth]{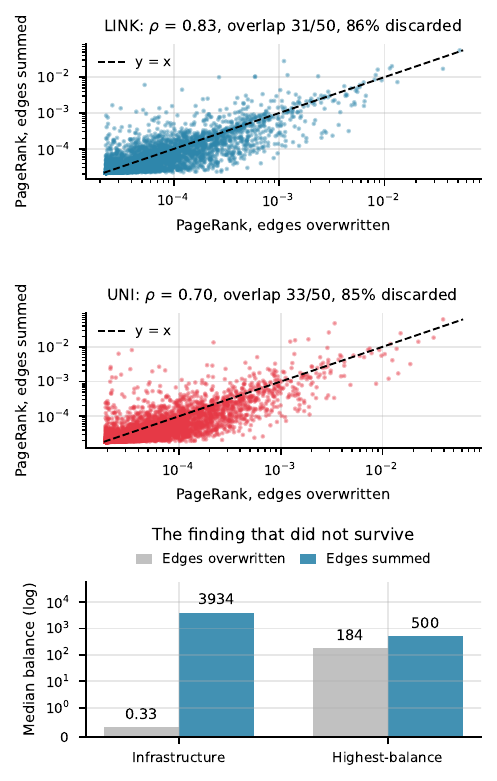}
  \caption{A one-line implementation choice and the finding it reversed.
  Left panels: PageRank computed with repeated transfers summed (vertical
  axis) against the same quantity with repeated transfers allowed to
  overwrite one another (horizontal), on the $10^4$ graph of each ledger.
  The node set, the edge set and the degree sequence are identical between
  the two; only the weights differ. Right panel: the median balance of the
  infrastructure cluster and of the highest-balance cluster on LINK under
  each construction. Under the faulty construction the infrastructure
  cluster appears to hold essentially nothing, which supported a
  cross-token contrast that the corrected construction removes.}
  \label{fig:edgeweight}
\end{figure}

We further document a case that is harder to dismiss as a matter of taste.
Repeated transfers between the same ordered pair of addresses dominate both
ledgers, and whether they are summed or allowed to overwrite one another is
a one-line implementation choice in standard graph libraries. Choosing the overwrite behaviour discards roughly $85\%$ of transferred
volume and overturns a finding (Fig.~\ref{fig:edgeweight})---that wealth and structural position are borne by disjoint
clusters---that had already been written into a draft of this paper
(Section~\ref{sec:edgeweight-case}). We report the correction, the
superseded result, and the archived artefacts of both.

We conclude that on-chain concentration should be reported as a specified
range rather than a point estimate, and we give the minimal specification
set required for two studies to be comparable
(\ref{app:sensitivity}).

\subsubsection{A Parameterised Criterion for Hidden Brokers (C2)}
We operationalise Burt's notion of brokerage \cite{burt1992structural} for
ledgers in which most participants carry no label. A \emph{hidden broker} is
an address that, under a structural centrality measure $s$ and a reference
label snapshot $L$, falls in the top percentile of $s$, holds no more than
the median balance, and is unlabelled in $L$. Stating $s$ and $L$ as explicit
parameters is the point: the criterion is sensitive to both, and we quantify
that sensitivity rather than suppressing it. Applied to our two ledgers it
recovers $30$ and $18$ addresses, twelve shared, of which six resolve to no
named protocol under manual on-chain verification.

A matched control experiment locates the source of that performance. Drawing
control addresses from the same eligibility pool and matching them on
counterparty count, we find that $35\%$ of controls resolve to named routing
protocols against $37\%$ of the selected brokers---a difference that is not
detectable at our sample size. Against an unmatched control the same figure
is $0\%$. The discriminative work is therefore done by the criterion and, in
particular, by counterparty count; the graph neural network we use to
instantiate $s$ recovers that signal without adding to it. We report this
negative result because it bears directly on how such findings should be
attributed.

\subsubsection{Three Measurement Layers and a Revocable One (C3)}
Ownership and routing answer different questions, and on these ledgers they
are close to statistically independent: across addresses with a positive
balance, rank correlations between balance and counterparty count are
$+0.081$ and $+0.032$, and between balance and betweenness $-0.028$ and
$-0.051$. That independence is the mechanism permitting a Gini of $0.99$ to
coexist with an unconcentrated clearing layer, and it means neither layer's
verdict can be inferred from the other's.

Neither layer, however, reaches the authority to change the rules under which
routing occurs. On the governance ledger we add a rule layer measured by the
Nakamoto coefficient and find it almost disjoint from the routing layer:
among the fifty largest senders by weekly outflow, five hold enough UNI to
open a governance proposal, and $71\%$ of the proposal-eligible addresses
transacted not at all during the window. Two voting-capable addresses suffice
to reach quorum. Whatever contestability the routing layer exhibits is held
at the pleasure of a rule layer that does not operate it.

\subsection{Decentralization and Inequality Measurement}
Quantifying decentralization has been a central theme in Web3 literature. Early technical metrics focused on consensus-layer topology, employing measures such as the Nakamoto Coefficient \cite{grajales2022measuring} and Shannon Entropy to quantify mining pool concentration.

As smart contract platforms matured, scholarly attention expanded to the application and token ledger layers. Recent empirical studies \cite{gudgeon2020defi, campajola2023evolution, zarir2021developing} have applied economic wealth concentration metrics—most notably the Gini Coefficient and Top-$k$ holding percentages—to assess token distribution. These studies report inequality, with Gini scores frequently exceeding 0.90 across major DeFi tokens.

A parallel line of work argues that decentralization is not a scalar but a property that must be measured separately at each layer of the stack. Ovezik et al.\ \cite{ovezik2022sok} formalise this as a stratified assessment, decomposing a blockchain system into hardware, software, network, consensus, tokenomics, client-API, governance and geography layers and showing that a system may be well distributed in one and concentrated in another. Our two-scale ownership/routing decomposition and the rule layer we add in Section~\ref{sec:governance} belong to that family; what we contribute is a clearing-and-routing layer, which that taxonomy does not isolate, together with a representation-learning instrument for measuring it.

A recurring limitation is the treatment of token ledgers as homogenous sovereign economies. By benchmarking token Gini coefficients exclusively against national household wealth statistics \cite{ubs2024wealth}, prior work overlooks the functional nature of utility tokens. Previous studies also neglect industrial market concentration metrics, such as the Herfindahl--Hirschman Index (HHI), which are standard in antitrust analysis \cite{doj2010horizontal}. We address that gap by evaluating token ledgers against their functional counterparts.

\subsection{Social Capital, Structural Holes, and Positional Rent Extraction}
To explain how power operates independently of capital accumulation on-chain, we draw upon Ronald Burt's theory of Structural Holes \cite{burt1992structural, burt2004structural}. In social capital theory, a structural hole exists when there is a gap or lack of direct connection between two distinct clusters within a network. An actor who bridges a structural hole gains positional advantage, acting as an indispensable broker who controls information flow and extracts positional rents without necessarily possessing intrinsic capital \cite{granovetter1973strength, burt2004structural}.

In traditional corporate and trade networks, structural hole brokers leverage social trust and exclusive relationships. In smart contract ecosystems, however, structural holes emerge naturally between fragmented liquidity pools, isolated Decentralized Exchanges (DEXs), and user execution interfaces.

The present work bridges social network theory and decentralized finance by showing that on-chain ``hidden brokers'' (e.g., DEX aggregators and solvers) function as algorithmic embodiments of Burt's structural hole bridges. These smart contracts maintain zero token balances ($Balance \approx 0$), yet command substantial topological control by bridging disconnected liquidity pools. They extract non-custodial transactional rents (the ``Tollbooth Economy'') purely through their strategic network positioning, suggesting that structural power in Web3 can be largely detached from static wealth ownership.

\subsection{Historical and Methodological Context}
The observation that removing a central issuer does not remove
intermediation has a long history. Medieval merchant banking concentrated
clearing in a handful of houses \cite{deroover1948medici}, and the
nineteenth-century U.S.\ clearinghouses coordinated liquidity among nominally
independent banks without any statutory monopoly \cite{gorton1987joint,
kindleberger2011manias}. The same dynamic is visible on public ledgers once
one looks past ownership to routing. Graph neural networks \cite{kipf2017semi,hamilton2017inductive}
and graph-based forensics on blockchain transaction networks
\cite{weber2019anti} supply the instrument we use to operationalise
structural position; we treat the choice of representation as a parameter
rather than a contribution, for reasons Section~\ref{sec:control} makes
explicit.

\section{Data and Graph Construction}
\label{sec:data}

We study two ERC-20 transfer networks selected to be functionally
dissimilar. The primary subject is the Chainlink token (LINK, contract
\texttt{0x5149\ldots f986ca}), an oracle-infrastructure asset and one of the
most actively traded ERC-20 tokens on Ethereum. The comparison subject is
the Uniswap governance token (UNI, contract
\texttt{0x1f98\ldots 01f984}), whose design premise is broadly distributed
community control. Contrasting an infrastructure token with a governance
token tests whether the concentration patterns we document are specific to
a protocol's function or general to permissionless ledgers.

Transfer records and balances for both tokens are extracted from the public
Ethereum ledger via Google BigQuery using an identical query template, over
an \emph{identical} 90-day observation window (2026-03-26 to 2026-06-24).
Fixing the window is essential: it removes market-regime differences as a
confound, so any divergence between the two ledgers is attributable to the
asset rather than to the period. Within this window LINK records
$1{,}233{,}497$ transfers among $642{,}126$ active addresses, and UNI
records $433{,}856$ transfers among $38{,}667$ active addresses.

Balances are computed as cumulative net flow over each token's full
transfer history rather than over the observation window alone, yielding
$885{,}011$ LINK addresses and $386{,}029$ UNI addresses with strictly
positive balance. All distributional statistics reported in this paper are
computed over these positive-balance sets; addresses that have interacted
with the token but hold a zero net position are excluded, since they
contribute no mass to either the Gini coefficient or the HHI.

\paragraph{Mint, burn, and zero-value records}
The BigQuery \texttt{token\_transfers} table records issuance and
destruction as ordinary transfers involving the null address or the
conventional burn address. These are protocol events rather than market
activity, and we remove them from \emph{both} endpoints before graph
construction: $3$ such records for LINK and $1{,}498$ for UNI. Zero-value
transfers ($26{,}311$ for LINK, $2{,}376$ for UNI) carry no mass and affect
no concentration measure, but they do enlarge the set of addresses counted
as active; we retain them in the main specification.

\paragraph{Edge construction and weight aggregation}
We model the ledger as a directed, weighted graph $G = (V, E)$, where each
node $v \in V$ is an address and each edge $(u, v) \in E$ summarises all
transfers from $u$ to $v$ within the window. The edge weight is the
\emph{sum} of the transferred amounts:
\begin{equation}
w(u,v) \;=\; \sum_{t \,\in\, \mathcal{T}(u,v)} \mathrm{amount}(t),
\label{eq:edgeweight}
\end{equation}
where $\mathcal{T}(u,v)$ is the set of transfers from $u$ to $v$ in the
window. This aggregation is not incidental. Repeated transfers between the
same ordered pair are the dominant mode of activity on both ledgers:
$36.7\%$ of LINK records and $83.9\%$ of UNI records are repeat edges across
the full ledger, rising to $90.9\%$ and $91.7\%$ within the activity-ranked
core subgraphs defined below. Treating such an edge as a single transfer
rather than as their sum would discard $85.6\%$ (LINK) and $85.3\%$ (UNI) of
transferred volume in the core graphs. The prevalence of repeat interaction
is itself informative: it is the signature of a routing layer in which a
small number of contracts intermediate the same counterparties week after
week, a pattern we return to in Section~\ref{sec:edgeweight-case}. To assess
robustness across orders of magnitude, we construct three nested subgraphs
by activity, containing $9{,}876$ ($\sim\!10^4$), $49{,}587$
($\sim\!5\times10^4$), and $641{,}276$ ($\sim\!6.4\times10^5$) nodes,
respectively. The largest graph is extremely sparse, with a mean degree of
approximately $1.2$---a structural property that, as shown later, causes
classical betweenness centrality to degrade. Because the UNI network is an
order of magnitude smaller, the cross-token structural comparison
(Section~\ref{sec:comparative}) is conducted at the matched $10^4$ scale,
where both graphs are directly comparable in size.

\subsection{Node Features}
\label{sec:features}
For each node we compute eight structural features: in-degree, out-degree,
weighted in-degree, weighted out-degree, the number of distinct
counterparties, PageRank, approximate betweenness centrality, and token
balance. Heavy-tailed quantities are stabilized with a $\log(1+x)$
transform followed by $z$-score standardization. Approximate betweenness is
computed by sampling ($k = 500$ pivots for the $10^4$ and $5\times10^4$
graphs; $k = 5{,}000$ for the $6.4\times10^5$ graph).

A central methodological safeguard concerns \emph{feature leakage}. Because
token balance is used as a prediction target in part of our evaluation
(Section~\ref{sec:c1}), balance is \emph{excluded} from the features fed to
the GNN whenever it serves as a label. The GNN therefore operates on the
seven remaining structural features, ensuring that no reported result is
contaminated by trivially predicting a target from itself.

\section{Method}
\label{sec:method}

\subsection{GraphSAGE Embedding}
\label{sec:sage}
We learn unsupervised node embeddings with a two-layer
GraphSAGE~\cite{hamilton2017inductive} encoder. The representation of node
$v$ at layer $k$ is updated by aggregating its neighborhood:
\begin{equation}
\label{eq:sage}
\mathbf{h}_v^{(k)} = \sigma\!\left(\mathbf{W}^{(k)} \cdot
\mathrm{CONCAT}\!\left(\mathbf{h}_v^{(k-1)},\;
\mathrm{MEAN}_{u \in \mathcal{N}(v)}\, \mathbf{h}_u^{(k-1)}\right)\right),
\end{equation}
where $\mathcal{N}(v)$ is the set of neighbors of $v$, $\mathbf{W}^{(k)}$ is
the learnable weight matrix at layer $k$, and $\sigma$ is a nonlinear
activation. The encoder maps the seven-dimensional input features through a
hidden layer of width $32$ to a final embedding of dimension $16$.

Neighbourhood aggregation of this form is one instance of graph propagation,
and its cost is the binding constraint on how far the analysis scales. The
same constraint applies to the hand-crafted features it is compared against:
exact betweenness is infeasible at our largest scale, and the sampled
estimator we substitute degenerates on a graph of mean degree $1.2$
(Section~\ref{sec:limitations}). Approximate propagation
algorithms~\cite{wang2021agp} address this class of problem directly, and we
return to them in Section~\ref{sec:future} as the route by which the
framework might be applied to a full ledger rather than to activity-ranked
subgraphs of it.

The model is trained for $200$ epochs with the Adam optimizer (learning
rate $0.01$) under an unsupervised objective that encourages the embeddings
of transacting pairs (positive edges) to be similar while pushing apart
randomly sampled negative pairs. Training is performed on Apple MPS
hardware without CUDA; all embeddings are computed inductively, so the same
encoder generalizes across the three graph scales.

\subsection{Evaluation Protocol for Structural Power}
\label{sec:measures}
To test whether the learned embeddings capture structural power beyond
classical descriptors, we evaluate three feature sets---classical
centrality (PageRank and betweenness), the full hand-crafted structural
features, and the unsupervised GNN embedding---against three independent
targets:
\begin{enumerate}
    \item \textbf{External label (\texttt{is\_core}).} Predicting an
    Etherscan-derived entity label that, by construction, never enters the
    model input; the number of positive nodes is $74$, $86$, and $161$ at
    the three scales.
    \item \textbf{Top-balance nodes.} Predicting membership in the top 5\%
    by balance, with balance excluded from the GNN input.
    \item \textbf{Neighborhood enrichment.} Measuring how strongly each
    representation concentrates same-class (core) neighbors in embedding
    space, relative to a random baseline.
\end{enumerate}
All classifiers are evaluated with stratified five-fold cross-validation
(shuffled, fixed seed $42$); reported AUC values are the mean across folds
with the corresponding standard deviation. Confidence intervals, where
given, are obtained by bootstrap resampling of the predictions (fit-once,
$B = 1000$).

\section{Experiments}
\label{sec:exp}

The empirical material follows the evaluation protocol of
Section~\ref{sec:measures}. Sections~\ref{sec:c1} and~\ref{sec:ablation}
establish the instrument, Section~\ref{sec:c2} traces concentration over
time, Section~\ref{sec:comparative} replicates the analysis on UNI,
Section~\ref{sec:c3} defines and applies the hidden-broker criterion,
with the matched control in Section~\ref{sec:control}, and
Section~\ref{sec:w7} reports a stress event. The specification analysis
that constitutes our primary contribution is in Section~\ref{sec:hhi}.

Unless otherwise stated, all classifiers are evaluated with stratified
five-fold cross-validation (shuffled, fixed seed), and reported AUC values
are the mean across folds with the corresponding standard deviation.
Confidence intervals, where given, are obtained by bootstrap resampling
of the predictions ($B=1000$). Crucially, node balance is \emph{excluded}
from the GNN input features whenever balance is used as a prediction target,
so that no evaluation is subject to feature leakage.

\subsection{The Instrument: Recovering Core Infrastructure from Structure
Alone}
\label{sec:c1}

Later sections require an operationalisation of ``structurally central''
that works on a ledger where most addresses carry no label. This section
establishes one and, equally importantly, establishes its limits: classical
centrality fails at this task by a wide margin, learned representations
succeed, and the several learned representations we tested are not
distinguishable from one another or from a well-specified hand-crafted
feature set. We therefore treat the choice of representation as a parameter
of the criterion rather than as a contribution, and we quantify what that
parameter is worth. We use two complementary lines of evidence, each at
three graph scales (approximately $10^4$, $5\times10^4$, and $6.4\times10^5$
nodes), and compare three feature sets: classical centrality (PageRank and
betweenness), the full set of hand-crafted structural features, and the
unsupervised GNN embedding.

\paragraph{Evidence 1: predicting an external label}
The most stringent test predicts the external \texttt{is\_core} label
(derived from Etherscan entity tags), which by construction can never enter
the model input and is therefore immune to leakage. As shown in
Table~\ref{tab:c1_iscore}, classical centrality performs barely above
chance at every scale, while both the hand-crafted structural features and
the GNN embedding recover the label well at the two smaller scales. At the
$10^4$ scale the GNN reaches $0.912 \pm 0.021$ against $0.896$ for the full
structural features and $0.600$ for centrality.

The GNN's margin over the structural features is small and shrinks
monotonically with graph size: $+0.016$ at $10^4$, $+0.007$ at
$5\times10^4$, and $-0.002$ at $6.4\times10^5$. Measured against the
seed-level standard deviation, none of these differences exceeds one
standard deviation, and we do not claim an AUC advantage for the GNN over a
well-specified structural feature set. What the table does establish, and
establishes decisively, is the failure of classical centrality: a gap of
roughly $0.30$ AUC at the smaller scales that no amount of seed variation
closes. The degradation of both learned and hand-crafted representations at
$6.4\times10^5$ has a simple cause---the positive rate falls from $0.75\%$
to $0.025\%$ ($161$ anchors among $641{,}276$ nodes), and every method
suffers under that imbalance.

\begin{table}[H]
\centering\small
\caption{five-fold cross-validated AUC for predicting the
external \texttt{is\_core} label. The label is never part of the model
input, ruling out feature leakage. Centrality and structural baselines are
deterministic given the graph; the GNN figure is the mean over independent
random initialisations ($10$ seeds at $10^4$, $5$ at $5\times10^4$, $3$ at
$6.4\times10^5$) with the standard deviation \emph{across seeds}. Bold marks
a result whose advantage exceeds one seed-level standard deviation.}
\label{tab:c1_iscore}
\begin{tabular}{lccc}
\toprule
Scale & Centrality & Full struct.\ features & \textbf{GNN} \\
\midrule
$10^4$            & $0.600$ & $0.896$ & $\mathbf{0.912 \pm 0.021}$ \\
$5\times10^4$     & $0.657$ & $0.914$ & $\mathbf{0.921 \pm 0.011}$ \\
$6.4\times10^5$   & $0.656$ & $0.759$ & $0.757 \pm 0.026$ \\
\bottomrule
\end{tabular}
\end{table}

As a secondary check, the representation also recovers top-balance addresses
without being given balance as input; we report it in
\ref{app:instrument} because it bears on none of the three
contributions directly.

\paragraph{Evidence 2: neighborhood enrichment}
Finally, we measure how strongly each representation concentrates
same-class (core) neighbours in embedding space. This is where the GNN's
contribution actually lies, and it is stable in exactly the way the AUC
margin is not: the GNN attains between $2.48$ and $2.69$ times the
neighbourhood purity of the dimension-matched structural features at every
scale (Table~\ref{tab:c1_enrich}), across a $65$-fold range in node count.

We report absolute purity rather than fold-enrichment over the random
baseline, and the distinction is not cosmetic. Because the random baseline
is the positive rate, it falls by a factor of $30$ from $10^4$ to
$6.4\times10^5$, so enrichment multiples inflate mechanically with graph
size. Classical centrality is the clearest case: its enrichment rises from
$2.0\times$ to $12.4\times$ across the three scales, which reads as
improvement, while its absolute purity \emph{falls} from $0.0149$ to
$0.0031$. At the largest scale, only three in a thousand of the neighbours
that centrality places next to a core node are themselves core. Any
enrichment figure reported on graphs of differing size should be
accompanied by the underlying purity, and we do so throughout.

\begin{table}[H]
\centering\small
\caption{absolute neighbourhood purity of core nodes
(fraction of a node's $k=10$ nearest neighbours in embedding space that are
themselves core). We report purity rather than fold-enrichment because the
random baseline shrinks with graph size ($0.0075$, $0.0017$, $0.00025$ at
the three scales), so an enrichment multiple can rise while absolute
performance falls. The final column is the GNN-to-structural purity ratio,
which is invariant to that baseline and is therefore comparable across
scales.}
\label{tab:c1_enrich}
\setlength{\tabcolsep}{4pt}
\begin{tabular}{@{}lcccc@{}}
\toprule
      & Central- & Struct.  &              & GNN /   \\
Scale & ity      & features & \textbf{GNN} & struct. \\
\midrule
$10^4$          & $0.0149$ & $0.0959$ & $\mathbf{0.2578}$ & $2.69\times$ \\
$5\times10^4$   & $0.0081$ & $0.0988$ & $\mathbf{0.2607}$ & $2.64\times$ \\
$6.4\times10^5$ & $0.0031$ & $0.0516$ & $\mathbf{0.1282}$ & $2.48\times$ \\
\bottomrule
\end{tabular}
\end{table}

\paragraph{What scales and what does not}
Fig.~\ref{fig:c1_iscore} sets the two evidence lines side by side, and
reading them together yields a sharper claim than any of them alone. The GNN's AUC advantage over hand-crafted structural
features does \emph{not} survive scaling: it falls from $+0.016$ to
$-0.002$ and is within seed noise throughout. Its neighbourhood-purity
advantage does: $2.69\times$, $2.64\times$, $2.48\times$ across a $65$-fold
range in node count. The two facts are consistent if the GNN's contribution
is to the \emph{geometry} of the embedding rather than to the linear
separability of the label---it places structurally similar addresses near
one another, which is what Section~\ref{sec:c3} relies on, without
necessarily making them easier to rank with a linear probe.

Two degradations are worth recording as properties of the data rather than
of any method. First, absolute purity for the learned and hand-crafted
representations roughly halves at $6.4\times10^5$---while centrality's falls
further still, to $38\%$ of its $5\times10^4$ value---tracking a positive rate
that drops to $0.025\%$.
Second, approximate betweenness degenerates on the largest, extremely
sparse graph (mean degree $\approx 1.2$), where it predicts
\texttt{is\_core} at chance level. Neither is an implementation artifact,
and the second reinforces the broader claim that hand-crafted centrality is
insufficient to characterise structural power at the scales at which these
ledgers actually operate.

\subsection{Ablation and Baseline Comparison}
\label{sec:ablation}

\begin{table}[H]
\centering
\caption{Ablation on the $10^4$ LINK graph. Learned representations---the GNN
and node2vec alike---separate core infrastructure far better than classical
centrality. The constant-input row measures a degeneracy of mean aggregation
under uninformative features, not the information content of topology;
node2vec is the appropriate topology-only control. Figures are means over
random initialisations with seed-level standard deviations.}
\label{tab:ablation}
\begin{tabular}{lc}
\toprule
Configuration & \texttt{is\_core} AUC \\
\midrule
GNN (full features)        & $\mathbf{0.912 \pm 0.021}$ \\
node2vec (topology only)   & $\mathbf{0.914 \pm 0.011}$ \\
GNN (constant input)       & $0.461 \pm 0.051$ \\
Classical centrality       & $0.600$ \\
\bottomrule
\end{tabular}
\end{table}

Three controls bound what the instrument of Section~\ref{sec:c1} is doing.

\paragraph{Learned representations versus hand-crafted centrality}
node2vec, a purely topological embedding run under the same multi-seed
protocol, is statistically indistinguishable from the GNN on LINK
($0.912 \pm 0.021$ against $0.914 \pm 0.011$; pooled $z = -0.21$) and behind
it on UNI ($0.846 \pm 0.029$ against $0.800 \pm 0.023$; $z = +3.31$). The
decisive gap is therefore between learned representations of any kind and
classical centrality ($0.600$), not between architectures, and the GNN's
edge over node2vec is token-dependent.

\paragraph{The constant-input ablation, and why it proves less than it
appears} Replacing all node features with a constant collapses
\texttt{is\_core} AUC to $0.461 \pm 0.051$. We caution against the natural
reading. With constant inputs, mean aggregation in GraphSAGE returns a
near-identical vector for every node, so the ablation measures a degeneracy
of that architecture under degenerate input rather than the information
content of topology. node2vec, which uses topology alone, reaches $0.914$ on
the same graph. Topology is highly informative; this particular
architecture cannot extract it without features to aggregate.

\paragraph{Two representations, two strengths}
node2vec attains \emph{higher} neighbourhood purity than the GNN on both
ledgers ($0.293$ against $0.258$; $0.170$ against $0.111$), which appears to
contradict the broker results until one examines the composition of each
method's top percentile. Among the top $1\%$ by score, $34\%$ carry a public
label under node2vec against $16\%$ under the GNN on LINK ($21\%$ against
$14\%$ on UNI). node2vec ranks already-catalogued infrastructure highest,
which maximises purity measured against labelled anchors; the GNN promotes
unlabelled contracts, which is what survives the no-label filter defining a
hidden broker. One method recalls what is known, the other surfaces what is
not---and Section~\ref{sec:control} shows that neither advantage is large
enough to attribute the broker findings to the representation.

Table~\ref{tab:ablation} reports these configurations; feature-group
ablations, the scale evidence and a graphical version
(Fig.~\ref{fig:ablation}) are in \ref{app:instrument}.

\subsection{Temporal Stability of Extreme Concentration}
\label{sec:c2}

\paragraph{A stable, extreme inequality}
Aggregating transfers into weekly windows, we find that concentration is
both extreme and remarkably stable over the 90-day period. On LINK the
weekly flow Gini averages $0.979$ (range $[0.972, 0.989]$), the top 1\% of
senders account for $81.1\%$ of weekly outflow, and the top 50 senders hold
an average $70.8\%$ flow share week over week. Yet week-over-week retention
of that top 50 is only $56\%$, revealing a regime of ``stable membership,
rotating ranks'': the boundary of the elite is locked while internal
positions churn. UNI shows the same regime with a firmer boundary---flow
Gini $0.969$, top-1\% share $70.7\%$, top-50 flow share $77.4\%$, and
retention $63\%$.

\begin{figure*}[t]
  \centering
  \includegraphics[width=0.95\linewidth]{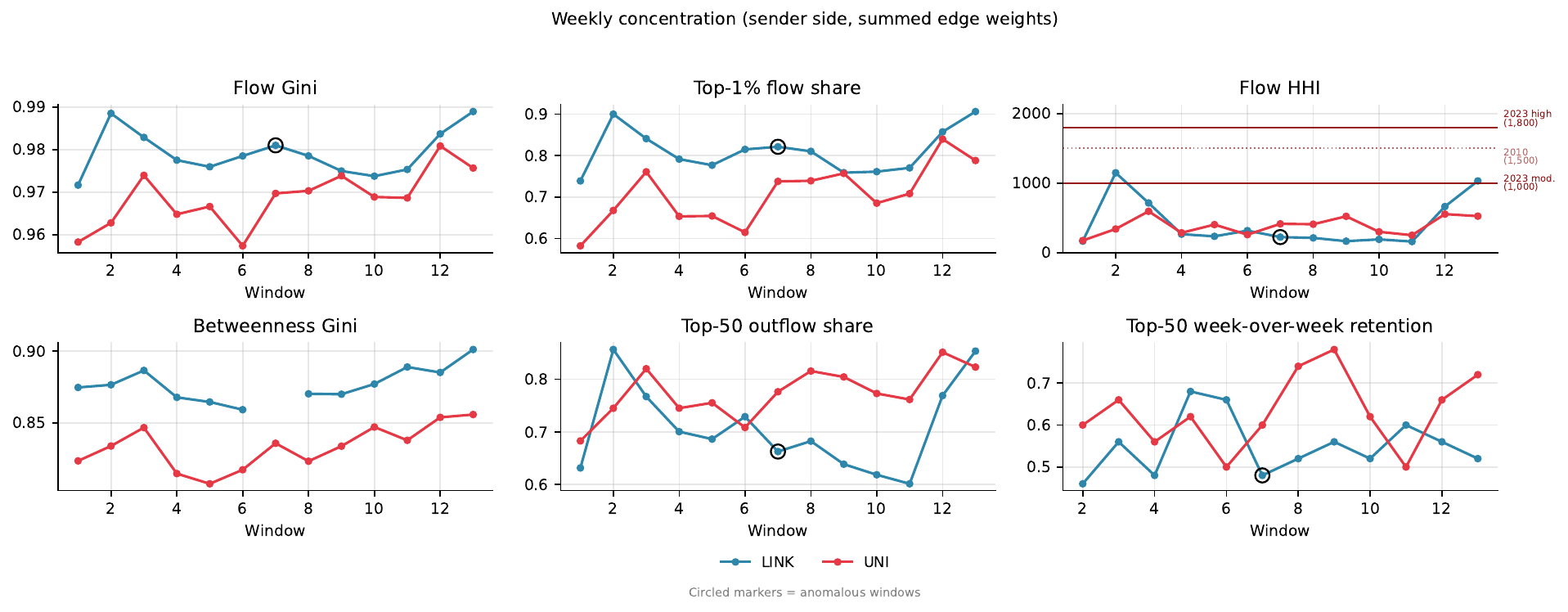}
  \caption{Weekly concentration dynamics for both ledgers over the same
  90-day window: flow Gini, top-1\% flow share, flow HHI against the DOJ
  threshold, betweenness Gini, top-50 flow share, and week-over-week
  retention of the top 50. All quantities are sender-side and computed on
  the aggregated edge weights of Equation~\ref{eq:edgeweight}. Concentration
  is high and trendless. Every window is unconcentrated under the 2010
  thresholds; under those of 2023, LINK's peak week crosses into the moderate
  band while UNI's does not. The circled point is LINK's coordinated
  phishing-airdrop window, which the outlier detector flags and which has no
  counterpart in the UNI series; its betweenness Gini is omitted because
  pivot sampling degenerates at $550{,}233$ nodes.}
  \label{fig:c2}
\end{figure*}

Fig.~\ref{fig:c2} shows all six weekly series for both ledgers. All weekly
quantities in this section are computed on the sender side and on the
aggregated edge weights of Equation~\ref{eq:edgeweight}. We report
retention only for window pairs in which neither window is anomalous, since
the week-7 event described in Section~\ref{sec:w7} contaminates both its
own retention value and that of the following week.

\paragraph{Window-wise re-training with embedding alignment}
To check that this picture is not an artefact of reusing one global
embedding, we re-train a separate GraphSAGE encoder on each weekly window and
align consecutive windows by orthogonal Procrustes. The anomalous window~7 on
LINK is excluded: at $550{,}233$ nodes it is more than thirty times the size
of a typical window and reflects a one-off event rather than ordinary routing
(Section~\ref{sec:w7}). Alignment
results for both ledgers are reported together in
Section~\ref{sec:comparative}.

\subsection{Cross-Token Validation: LINK versus UNI}
\label{sec:comparative}

A single-asset study cannot distinguish a property of the protocol from a
property of permissionless ledgers in general. We therefore replicated the
concentration analysis on a second, functionally dissimilar asset---the
Uniswap (UNI) governance token---over the \emph{identical} 90-day window
(2026-03-26 to 2026-06-24), using an identical extraction query and an
identical measurement script. Holding the observation period fixed removes
market-regime effects as a confound, so any difference between the two
tokens is attributable to the asset rather than to the period.

The comparison is deliberately adversarial to our own thesis. UNI is a
governance token whose entire design premise is decentralized, broadly
distributed control; if concentration were an artifact of LINK's
infrastructural role, UNI should look markedly more egalitarian. It does
not. UNI is \emph{more} unequal on every distributional measure: a balance
Gini of $0.998$ against $0.990$, a top 1\% share of $98.4\%$ against
$93.2\%$, and a top 10 share of $51.7\%$ against $32.9\%$. A token marketed
on the promise of community governance concentrates ownership more tightly
than the oracle network it helps price.

At the market-structure scale the two assets converge rather than diverge.
Mean weekly routing concentration is $421$ for LINK and $386$ for UNI, both
inside the unconcentrated band on the mean; LINK's peak week of $1{,}147$
enters the moderate band under the 2023 thresholds, while UNI's maximum of
$593$ stays below it on either convention. The remaining difference lies in balance HHI ($135$ for
LINK versus $872$ for UNI), and it too dissolves under scrutiny: protocol
holdings account for $21.0\%$ of observed LINK supply and $28.0\%$ of UNI
supply, with a further $10.8\%$ of UNI held at the burn address. Excluding
burn and issuer addresses brings the two ledgers to $115$ and $109$, a
difference of $6\%$. What looked like a six-fold gap in ownership
concentration is, on inspection, a difference in how much of its own supply
each protocol holds. We interpret these results in
Section~\ref{sec:discussion}.

\paragraph{The measurement instrument generalizes}
Distributional statistics alone would not establish that our \emph{method}
transfers across assets. We therefore re-ran the full instrument pipeline on UNI at
the matched $10^4$ scale, using an identical script, identical
hyperparameters, and an identical evaluation protocol. The two graphs are
almost exactly the same size ($9{,}765$ nodes for UNI against $9{,}876$ for
LINK), so no difference can be attributed to graph scale.

Table~\ref{tab:c1_cross} reports the outcome.
The GNN retains a decisive advantage over classical centrality on the
leakage-proof \texttt{is\_core} target for both assets: $0.912$ versus
$0.600$ on LINK, and $0.846$ versus $0.612$ on UNI. The most striking
regularity is not the GNN scores but the \emph{baseline} ones. Classical
centrality lands at $0.600$ and $0.612$ on two functionally unrelated
tokens---a difference of $0.012$. PageRank and betweenness are therefore
not failing to identify core infrastructure because of some peculiarity of
the Chainlink network; they plateau just above chance wherever we look,
which is what one expects if hand-crafted centrality simply does not encode
the property that distinguishes infrastructure from ordinary accounts. The
purity results tell the same story: the GNN sustains $2.69\times$ (LINK)
and $2.17\times$ (UNI) the neighbourhood purity of the dimension-matched
structural-feature baseline.

\begin{table}[H]
\centering\small
\caption{The instrument replicated across tokens at the matched $10^4$
scale. Identical pipeline, hyperparameters, and evaluation protocol.
Classical centrality plateaus near $0.61$ on both assets---a difference of
$0.012$ between two functionally unrelated ledgers---while both learned
representations retain a large margin on each. Neighbourhood purity is
reported rather than fold-enrichment, following the convention set in
Section~\ref{sec:c1}; the random baseline is the positive rate.}
\label{tab:c1_cross}
\begin{tabular}{llcc}
\toprule
Evidence & Representation & LINK & UNI \\
\midrule
\multirow{3}{*}{\texttt{is\_core} AUC}
 & Centrality             & 0.600 & 0.612 \\
 & Full struct.\ features & 0.896 & 0.830 \\
 & \textbf{GNN}           & \textbf{0.912} & \textbf{0.846} \\
\midrule
\multirow{3}{*}{Purity}
 & Centrality             & $0.015$ & $0.016$ \\
 & Full struct.\ features & $0.096$ & $0.051$ \\
 & \textbf{GNN}           & $\mathbf{0.258}$ & $\mathbf{0.111}$ \\
\addlinespace[2pt]
\multicolumn{2}{l}{\quad Random baseline} & $0.0075$ & $0.0058$ \\
\midrule
\multicolumn{2}{l}{Graph size / \texttt{is\_core} anchors}
 & 9{,}876 / 74 & 9{,}765 / 57 \\
\bottomrule
\end{tabular}
\end{table}

\paragraph{Temporal concentration is stable in both ledgers, but located
in different dimensions}
Applying the weekly windowing procedure to UNI yields thirteen windows and,
unlike LINK, no anomalous window: the automated detector (node count above
three times the median) flags nothing, confirming that the coordinated
airdrop of Section~\ref{sec:w7} is an event in the LINK series rather than a
general feature of token ledgers. Both series are stationary---the linear
trend in UNI's weekly flow Gini is $+0.0004$ per week, statistically
indistinguishable from flat, matching the absence of trend in LINK (Fig.~\ref{fig:c2}).

The two assets differ in \emph{where} their concentration sits. UNI is the
more unequal ledger by holdings (balance $Gini = 0.998$ against $0.990$) but
the less unequal by flow (mean weekly flow $Gini = 0.969$ against $0.979$;
top-1\% flow share $70.7\%$ against $81.1\%$). LINK inverts this. Structural
concentration, measured as the Gini of betweenness, is comparable
($0.833$ and $0.877$; the LINK figure averages the twelve windows in which
sampled betweenness is non-degenerate, excluding window~7).

The window-wise embedding analysis replicates as well. Re-training a
separate GraphSAGE encoder on each weekly window and aligning the resulting
spaces by orthogonal Procrustes succeeds for every retained pair on both
ledgers (UNI $12/12$, LINK $11/11$ after excluding the anomalous window~7),
with a minimum of $809$ common anchor nodes between adjacent windows on UNI
and $2{,}027$ on LINK. Of the addresses present in every window, $53$
persist throughout on UNI against $157$ on LINK. Normalised by the size of a
typical window's elite set these are $22\%$ and $18\%$---a closer agreement
between the two ledgers than the raw counts suggest, and one that holds
despite an order-of-magnitude difference in graph size.

The quantity of interest, mean inter-window drift of the persistent elite in
the aligned space, is $1.506$ on UNI against $1.370$ on LINK, a difference
of $0.136$ that is small relative to the spread of window-to-window drift
within either series. Two independently trained sequences of embeddings, on ledgers that
differ in function, size and holder composition, produce the same
characteristic rate at which structurally central addresses shift position.
Combined with the turnover result below, this makes the ``stable set,
rotating seats'' finding the most thoroughly replicated in the paper---it
holds under two measurement approaches (flow-share turnover and
embedding-space drift) on two assets.

The elite-turnover result replicates directly. Weekly retention of the
top-50 senders averages $63\%$ on UNI against $56\%$ on LINK, both computed
on decontaminated window pairs under the sender-side convention used
throughout. Roughly half the leading positions change hands from
one week to the next in both ledgers, while the aggregate concentration
they produce does not move. The ``stable set, rotating seats'' pattern is
therefore not an idiosyncrasy of the Chainlink network: extreme
concentration is reproduced week after week by a partially different cast
of intermediaries, which is what one expects when the routing layer is
contestable (Section~\ref{sec:c2}) but the barriers to reaching it are not.

We do not, however, report a cross-token comparison of the elite flow
\emph{share}. That statistic is defined over a fixed top-500 set, which
represents roughly $3\%$ of nodes in a typical LINK window but $10\%$ in the
smaller UNI windows ($4{,}041$--$6{,}182$ nodes against LINK's
$\sim\!16{,}000$); the two are not measuring the same object, and
re-normalising the cut would change the LINK figure reported earlier.

\paragraph{The role typology transfers, but one finding is
token-specific}
Re-running the clustering and broker-identification procedure on UNI
reproduces the overall architecture: a small, dense infrastructure cluster
($154$ nodes, $8.4\%$ core rate, median $78$ counterparties) sitting above a
broad periphery of retail addresses, closely mirroring LINK's ($169$ nodes,
$10.1\%$, median $80$). On both ledgers this cluster is simultaneously the
highest in core rate and in median balance---the infrastructure that clears
the ledger also holds the most of it.

An earlier construction of the edge weights produced the opposite pattern on
LINK; Section~\ref{sec:edgeweight-case} sets out why it did not survive.

The wealth--power divergence survives, but at the level where it is actually
measurable. Across individual nodes with a positive balance, the rank
correlation between balance and counterparty count is $+0.081$ on LINK and
$+0.032$ on UNI, and between balance and betweenness $-0.028$ and $-0.051$.
All four are indistinguishable from zero. Wealth and structural power are
not opposed on these ledgers; they are close to \emph{orthogonal}, and that
orthogonality is what allows a Gini of $0.99$ to coexist with an
unconcentrated routing layer. The cluster-level association reflects a
single small cluster of exchanges and routers that both hold and move
supply, and should not be read as a general coupling.

\begin{figure}[H]
  \centering
  \includegraphics[width=0.98\linewidth]{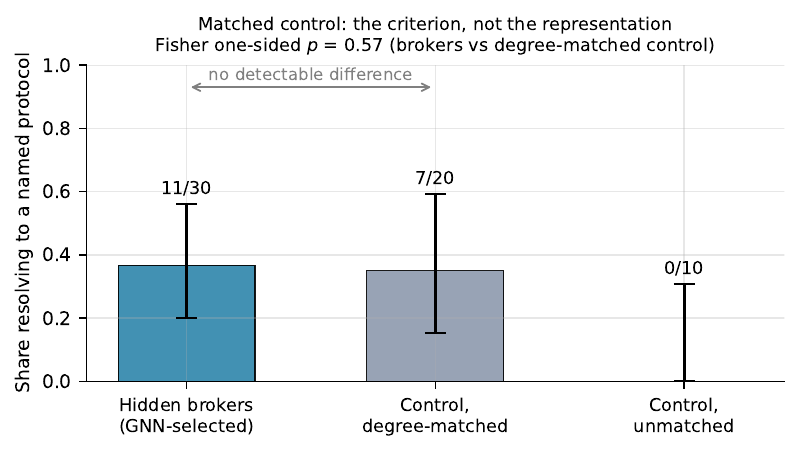}
  \caption{Matched control experiment on LINK. Bars give the share of each
  group resolving to a named routing protocol under blind manual
  verification, with Clopper--Pearson $95\%$ intervals. The degree-matched
  control is drawn from the same eligibility pool as the brokers and matched
  on counterparty count, so selection by the ensemble is the only systematic
  difference between the first two bars; no difference is detected. The
  unmatched control, drawn from the same pool without matching, resolves to
  no named protocol at all, which is what makes the matching necessary.}
  \label{fig:random_control}
\end{figure}

\subsubsection{A Matched Control: What the Criterion Does, and What the
Representation Adds}
\label{sec:control}
The brokers reported above---selected by the criterion set out in
Section~\ref{sec:c3}---were chosen by a GNN and then verified by hand.
That procedure cannot, on its own, establish that the selection was
informative: we inspected only the addresses the model chose. To test it we
drew a control group from the \emph{same} eligibility pool---unlabelled,
balance at or below the median, at least five counterparties, non-zero
sampled betweenness---restricted to addresses the ensemble did \emph{not}
select, and verified them under the same protocol.

Two features of the design are load-bearing. First, the control is matched
on counterparty count. The eligibility pool has a median of $7$ counterparties
against $59$ for the selected brokers, so an unmatched control would differ
from the brokers in degree as well as in selection, and any difference in
outcome could be attributed to either. Matching each broker to the eligible
non-broker closest in counterparty count brings the control's median to $53$,
after which selection is the only systematic difference remaining. Second,
verification was blind: the control addresses were shuffled together with a
subset of the brokers into a single list carrying no group labels, and the
group assignment was not consulted until verification was complete. The
blinded broker anchors resolved to named protocols at $50\%$ against a known
rate of $37\%$ for the full broker set, a difference small enough to indicate
that the standard of evidence did not shift under blinding.

The result does not favour the model. Of the twenty degree-matched controls,
seven ($35\%$) resolve to a named routing protocol, against eleven of thirty
($37\%$) among the selected brokers---a difference of $1.7$ percentage points,
with a one-sided Fisher exact $p = 0.57$ (Fig.~\ref{fig:random_control}).
Of the ten unmatched controls, drawn from the same pool without matching on
degree, \emph{none} resolves to a named protocol, and the difference between
the unmatched and matched controls is itself significant (one-sided $p = 0.038$; two-sided $p \approx 0.064$).

We read this as follows. The eligibility criterion---in particular the
counterparty threshold---is doing the discriminative work. Conditional on
degree, the GNN's ranking adds nothing detectable at this sample size. The
experiment is powered to exclude a difference of roughly thirty percentage
points but not a small one, so the honest statement is that no advantage was
detected rather than that none exists. Either way, the finding that these
addresses are real, uncatalogued routing infrastructure does not depend on
the representation used to surface them, and we have avoided attributing it
to one.

There is a corollary that strengthens rather than weakens the substantive
claim. Both groups are unlabelled in the Etherscan snapshot by construction,
yet roughly $36\%$ of \emph{both} resolve to named protocols under manual
verification. The label database is therefore missing about a third of the
high-degree routing infrastructure on this ledger, and that gap is a
property of the database, not of any method used to probe it.

\paragraph{Hidden brokers are a shared routing layer, not a token
artifact}
The broker sets overlap substantially across the two ledgers. Of the $18$
hidden brokers identified on UNI, $12$ are the same addresses recovered
among the $30$ on LINK, and the overlap concentrates at the top: seven of
the ten highest-counterparty UNI brokers also appear in the LINK set. Six
named routing contracts intermediate both assets: the Uniswap V4 Universal
Router, a second Universal Router deployment carrying cross-chain
extensions, Uniswap~X's Rizzolver solver, Bitget's DEX aggregator, LI.FI's
Diamond router, and a 0x~Protocol routing contract whose attribution rests
on verified source and deployer rather than on a public name tag. This
directly explains
the convergence documented in Section~\ref{sec:comparative}: the two tokens
exhibit near-identical routing concentration ($HHI_{\text{Flow}}$ of $421$
and $386$) because the routing is performed by a common, mutually competing
set of contracts rather than by token-specific infrastructure. Cross-token
replication thus does more than validate the method; it identifies the
object the method is measuring.

Two entries in that list qualify the claim in ways worth stating. 1inch
appears on both ledgers but through \emph{different} contracts---Aggregation
Router~V6 on LINK, Aggregation Executor~5 on UNI---so its cross-token
presence is a fact about the operator, not about any single address, which
is precisely the distinction Section~\ref{sec:entity} formalises. And Mayan's
Swift~v2 endpoint, which an earlier single-run analysis placed among the
shared brokers, falls just outside the top percentile on both ledgers under
the multi-seed ensemble (score percentiles $0.960$ and $0.917$); we remove
it from the list rather than retain a near-miss.

\begin{figure}[H]
  \centering
  \includegraphics[width=0.98\linewidth]{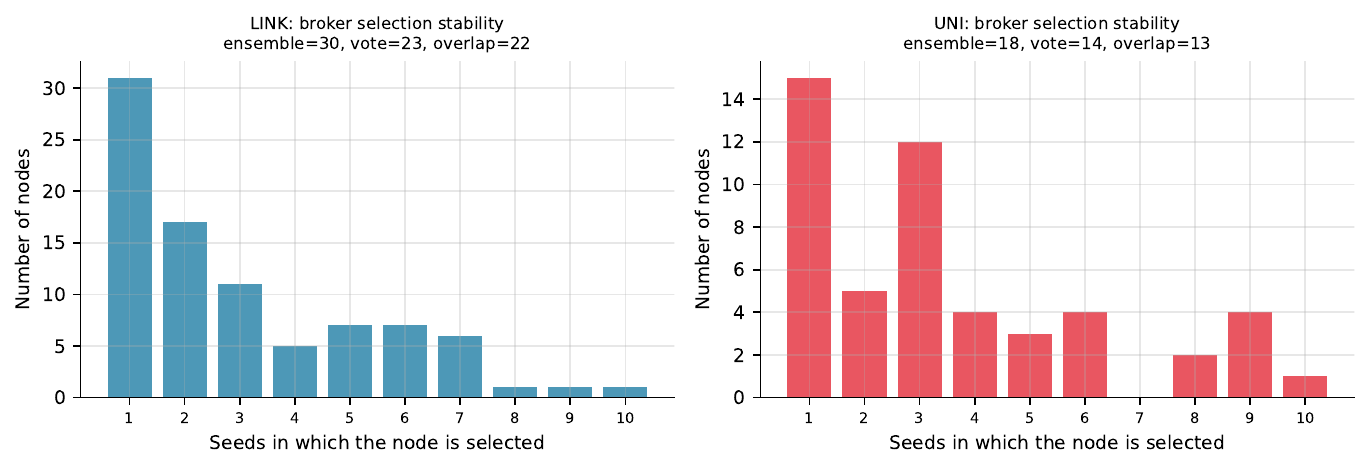}
  \caption{Selection frequency of hidden brokers across ten independent
  random initialisations. Most addresses that qualify in any run qualify in
  very few---$48$ of $87$ on LINK appear at most twice---which is the
  expected behaviour of a hard percentile cut near its boundary rather than
  a defect of the criterion. We therefore report a rank ensemble over the
  ten runs, together with the vote-consensus subset and their intersection,
  instead of the outcome of any single run.}
  \label{fig:broker_stability}
\end{figure}

\paragraph{Stability of the broker set}
\label{sec:brokerstability}
Because the criterion cuts a hard percentile, membership near the boundary is
sensitive to initialisation. Across ten runs on LINK, $87$ addresses are
selected at least once, $48$ of them no more than twice, and only two in nine
runs or more; the corresponding figures on UNI are $50$, $20$ and five
(Fig.~\ref{fig:broker_stability}). The
rank-ensemble specification we report is deterministic given a fixed set of
embeddings, and the intersection with the vote-consensus criterion---$22$
addresses on LINK and $13$ on UNI---provides a high-confidence subset whose
composition is stable under either rule. Where the text names an individual
contract, we do so only for members of that intersection.

\paragraph{Half of the shared brokers carry no verifiable identity}
Of the twelve brokers appearing on both ledgers, six resolve to a named
protocol under manual on-chain verification and six do not. The six
unidentified contracts are not obscure. Each records on the order of $10^4$
ERC-20 transfers, several remained active within minutes of inspection, and
their counterparties identify a specific and comparatively recent class of
infrastructure: intent-based solvers and aggregation relays, cross-chain
routing endpoints, and batch distribution contracts. What they lack is a
name tag.

One case makes the point sharply. The single highest-degree cross-token
broker ($230$ distinct counterparties, betweenness $0.011$, selected in nine
of ten seeds) carries no label of any kind, yet its deployer address also
deployed a contract that Etherscan labels ``1inch: Aggregation Executor 5''.
Two contracts from the same operator, one catalogued and one not, and the
uncatalogued one is the more structurally central of the pair. We record the
deployer relationship as evidence but do not assign the label, so that the
entity-resolution figures of Section~\ref{sec:entity} remain a conservative
lower bound.

That these contracts are absent from the label databases on which
conventional on-chain analysis depends is the substantive finding. A
structural method recovers them because it asks what a node \emph{does} in
the graph rather than what it is called---and the two questions have
measurably different answers.

\paragraph{Structural coreness is agnostic to legitimacy}
Manual verification of the brokers recovered only on UNI returned a more
heterogeneous population than on LINK. The highest-ranked of them is a
cross-chain router with $183$ counterparties; two others are
Etherscan-labelled MEV bots; and one is a contract that Etherscan flags as
\emph{compromised}, with an active vulnerability warning. Together with the
phishing-labelled address in the
LINK set, this establishes that the signature we exploit---high structural
coreness with zero custody---is shared by legitimate routers, extractive
automation and adversarial contracts alike. The method identifies
\emph{operational centrality}, not legitimacy. We regard this as a
qualification of the interpretation rather than a defect of the measurement:
a monitoring apparatus for systemic operational risk needs precisely a
signal that fires on whoever occupies the critical path, independent of
intent.

Three caveats bound this comparison. First, the GNN margin is smaller on
UNI than on LINK (\texttt{is\_core} $0.846$ against $0.912$; purity
$0.111$ against $0.258$), and UNI supplies fewer labelled anchors
($57$ against $74$), so its estimates carry more sampling noise. Second,
the full UNI transfer graph is an order of magnitude smaller than LINK's
($38{,}667$ against $642{,}126$ active addresses), which is why the
head-to-head comparison is conducted at $10^4$ rather than at full scale.
Third, two tokens remain a small sample: we regard this as evidence that
neither the concentration pattern nor the measurement instrument is
protocol-specific, not as proof of universality across the asset class.

\subsection{Applying the Hidden-Broker Criterion}
\label{sec:c3}

\paragraph{Eight structural roles}
Clustering the GNN embedding yields eight roles, which we name by jointly
anchoring on Etherscan labels and structural statistics
(Table~\ref{tab:typology}). Two findings stand out, and both replicate
across the two ledgers. First, the core infrastructure exhibits a
\emph{two-layer} structure: a small, dense cluster ($169$ nodes on LINK with
a $10.1\%$ core rate and a median of $80$ counterparties; $154$ nodes,
$8.4\%$ and $78$ on UNI) sits above a broad routing layer of one to two
thousand addresses with far lower core rates. Second, on both ledgers that
dense cluster is simultaneously the highest in core rate \emph{and} in median
balance ($3{,}934$ and $3{,}405$), so wealth and clearing function are not
borne by separate clusters. Cluster indices are k-means
labels and differ between the two ledgers; the correspondence is in the roles.
The one cluster with a non-trivial median balance outside the core
(cluster~1 on LINK, median $500$, core rate $0.37\%$) holds without routing,
but it accounts for $1{,}342$ of $9{,}876$ addresses and does not constitute
a separate stratum of comparable weight to the core.

\begin{table*}[t]
\centering\footnotesize
\caption{Role typology from clustering the GNN embedding at the matched
$10^4$ scale on both ledgers. Roles are named by combining label composition
with structural statistics; ``core rate'' is the fraction of
\texttt{is\_core} nodes and ``cp.'' the number of distinct counterparties.
The two ledgers are aligned by role rather than by cluster index, since the
indices $k$-means returns are arbitrary and carry no meaning across runs.
Sizes sum to $9{,}876$ and $9{,}765$, the full node count of each graph.}
\label{tab:typology}
\setlength{\tabcolsep}{4.5pt}
\renewcommand{\arraystretch}{1.0}
\begin{tabular}{@{}l rrrr c rrrr@{}}
\toprule
& \multicolumn{4}{c}{\textbf{LINK} ($9{,}876$ nodes)} & &
  \multicolumn{4}{c}{\textbf{UNI} ($9{,}765$ nodes)} \\
\cmidrule(r){2-5}\cmidrule(l){7-10}
Role & Size & Core rate & Med.\ cp. & Med.\ bal. & &
       Size & Core rate & Med.\ cp. & Med.\ bal. \\
\midrule
Core infrastructure    & $169$     & $10.06\%$ & $80$ & $3{,}934.45$ & &
                         $154$     & $8.44\%$  & $78$ & $3{,}404.89$ \\
Active routing layer   & $1{,}492$ & $3.28\%$  & $7$  & $0$          & &
                         $2{,}029$ & $1.53\%$  & $4$  & $0$          \\
Holders (low activity) & $1{,}342$ & $0.37\%$  & $2$  & $500.00$     & &
                         $839$     & $0.36\%$  & $3$  & $0$          \\
\addlinespace[2pt]
Retail / leaf          & $2{,}127$ & $0.09\%$  & $3$  & $0$          & &
                         $1{,}615$ & $0.25\%$  & $3$  & $0$          \\
Retail / leaf          & $2{,}090$ & $0.05\%$  & $3$  & $0$          & &
                         $2{,}575$ & $0.16\%$  & $2$  & $0$          \\
Retail / leaf          & $1{,}232$ & $0.00\%$  & $3$  & $0$          & &
                         $1{,}650$ & $0.12\%$  & $2$  & $0$          \\
Retail / leaf          & $810$     & $0.00\%$  & $3$  & $0$          & &
                         $537$     & $0.00\%$  & $3$  & $0$          \\
Retail / leaf          & $614$     & $0.00\%$  & $4$  & $0$          & &
                         $366$     & $0.00\%$  & $2$  & $0$          \\
\bottomrule
\end{tabular}
\end{table*}

\paragraph{Hidden brokers}
We define hidden brokers as nodes that are structurally core yet hold little
balance and carry no external label---power holders sitting on critical
paths without accumulating wealth. Operationally, a node qualifies if it
falls in the top $1\%$ of the \texttt{is\_core} classifier score, holds a
balance at or below the median, carries no external label, and additionally
transacts with at least five distinct counterparties and has non-zero
sampled betweenness. The last two conditions exclude degenerate cases: an
earlier specification without them admitted single-counterparty vanity
addresses that cannot, by construction, sit on any path.

Because a hard percentile cut is unstable under the stochasticity of GNN
training, we do not report a single run. We train ten embeddings from
independent initialisations, convert each classifier score to a percentile
rank, average the ranks, and cut the top $1\%$ once. This yields $30$
brokers on LINK and $18$ on UNI. A stricter alternative---requiring
selection in at least half of the individual runs---yields $23$ and $14$,
and the intersection of the two criteria yields $22$ and $13$. We report the
rank-ensemble set as the main specification and the intersection as a
high-confidence subset, and we give the full selection-frequency
distribution in Section~\ref{sec:brokerstability}, because the instability
is itself informative: of the $87$ LINK nodes selected in at least one run,
$48$ appear in no more than two.

Matched against a betweenness-based selection of the same size drawn from
the same eligibility pool, the two criteria agree on $7$ of $30$ nodes on
LINK ($23\%$) and $8$ of $18$ on UNI ($44\%$). Note that at equal set size
each method necessarily has as many unique selections as the other, so the
comparison licenses a statement about \emph{overlap} but not about one
method finding more brokers than the other.

We manually verified every address in the union of the two broker sets
through multi-source on-chain forensics (Etherscan name tags, verified
source code, and deployer attribution), $36$ addresses in total. Eleven of the $30$ LINK brokers and nine of the $18$ UNI brokers resolve to a
named protocol---DEX aggregators, swap routers, and algorithmic solvers---as
summarised in Table~\ref{tab:brokers}. Representative examples include the
1inch Aggregation Router~V6 and Aggregation Executor~5, MetaMask Swaps
Spender, the Uniswap~X (Rizzolver) solver, two distinct Universal Router
deployments, KyberSwap's Aggregator Executor, Velora's Augustus router, and
the OKX Labs DEX Router, together with two contracts attributable to the
0x~Protocol (verified source, deployer \texttt{deployer.zeroexprot...}, no
public name tag). All share the defining signature of a hidden broker: they
intermediate a very large number of counterparties---up to $313$ on a single
ledger and $230$ across both tokens---while holding zero token balance, forwarding routed value to off-chain treasuries
rather than accumulating it.

Counting per ledger, the remaining $28$ of the $48$ divide into two groups. Most route
substantial volume between DEX pools and aggregators but carry no name tag
and no verified source; we record them as unlabelled routing contracts
rather than attributing them, and several share a deployer with a contract
that Etherscan does label. Nine are high-frequency arbitrage
contracts---between $2.5\times10^4$ and $1.3\times10^5$ transactions each,
one deployed by an address Etherscan tags as an MEV builder---which we
label as MEV bots and, in the entity resolution of
Section~\ref{sec:entity}, deliberately leave unmerged, since two bots are
not one firm. We give counts rather than an exhaustive enumeration because
the distinction between the two groups rests on behavioural reading of the
transfer history, and the full per-address record is archived with the
replication package.

Two observations reinforce the value of the GNN. First, several of these
routers (e.g., 1inch~V6, Uniswap~X, Uniswap~V4) postdate the label snapshot
and are therefore \emph{absent from the external labels}, yet are recovered
purely from structure---direct evidence that the method surfaces real
infrastructure missed by label databases. Against the same eligibility
filter, an equally-sized node2vec selection recovers no named protocol at
all on either ledger, against $37\%$ on LINK and $50\%$ on UNI for the GNN; the two
embeddings rank different populations, and only one of them ranks unlabelled
infrastructure highly. Second, the procedure also
flagged one address subsequently tagged as a phishing contract
(\texttt{0xa819...bcdd}) and a small number of MEV bots; rather than
false positives, these illustrate that ``structural coreness without
custody'' is a signature shared by both legitimate routers and adversarial
automation---a distinction we return to in the Discussion. Verification must be conducted on the
token-transfer view rather than the default transaction view of a block
explorer, since contract-to-contract routers register few of the latter;
Section~\ref{sec:limitations} records the consequences of overlooking this.

\begin{table}[H]
\centering\small
\caption{Manually verified hidden brokers (subset, ranked by
counterparties). The GNN assigns all of them a core probability above
$0.98$ despite zero token balance; the majority are non-custodial DeFi
routing infrastructure.}
\label{tab:brokers}
\setlength{\tabcolsep}{4pt}
\begin{tabular}{@{}lrl@{}}
\toprule
Address & Cp. & Verified identity \\
\midrule
\texttt{0x7f54...be8a} & 313 & 0x Protocol Router \\
\texttt{0x74de...6631} & 277 & MetaMask: Swaps Spender \\
\texttt{0x1111...42a65} & 239 & 1inch: Aggregation Router V6 \\
\texttt{0x225a...dc17} & 177 & Rizzolver: Uniswap X \\
\texttt{0x6a00...1068} & 120 & Velora: Augustus V6.2 \\
\texttt{0x66a9...b8af} & 105 & Uniswap V4: Universal Router \\
\texttt{0x28b1...a183} & 91 & OKX Labs: DEX Router 2 \\
\texttt{0xbc1d...c973} & 48 & Bitget: DEX Aggregator \\
\bottomrule
\end{tabular}
\end{table}

\subsection{Case Study: A Coordinated Phishing Airdrop on the LINK Ledger}
\label{sec:w7}
One LINK window stands out: its transfer count is roughly twelve times the
series median. Thirty-eight verified contracts presenting themselves as
Chainlink giveaways account for at least $380{,}000$ of its $610{,}228$
transfers at $0.0014$ LINK each, and $97.5\%$ of the window's $549{,}223$
recipients receive exactly one transfer (\ref{app:phishing}).

The event is useful because it separates two statistics that are often
treated as interchangeable. Receiver-side flow Gini rises to $0.999$ and
recovers within a week; sender-side concentration and flow HHI barely move.
A flood of near-worthless transfers saturates an unweighted dispersion
measure while leaving value-weighted market structure untouched, which is a
concrete instance of the Gini--HHI divergence that Section~\ref{sec:hhi}
develops. Sampled betweenness degenerates on this window ($99.2\%$ of nodes
score zero at $550{,}233$ nodes), so we record its structural Gini as
missing rather than zero, and we exclude the window from the embedding
alignment of Section~\ref{sec:c2}.

\section{Discussion}
\label{sec:discussion}

This section discusses the economic, institutional and social implications of the empirical findings. The analysis proceeds in five steps: we first compare the wealth and transactional inequality of both ledgers against sovereign and asset-class benchmarks, then shift to a functional comparison using the Herfindahl--Hirschman Index and examine how far that comparison survives the discretionary choices it requires, interpret the GNN role typology, add the rule layer that neither ownership nor routing statistics can reach, and close with a re-examination of what ``decentralization'' means in practice.

\subsection{Wealth and Transactional Concentration}
Both ledgers are extraordinarily unequal in ownership: balance Ginis of
$0.990$ and $0.998$, with the top $1\%$ of addresses holding $93.2\%$ and
$98.4\%$ of supply. These exceed U.S.\ household wealth ($Gini \approx 0.75$)
\cite{ubs2024wealth} and sovereign official gold reserves ($\approx 0.85$--$0.90$
across countries, excluding private and jewellery holdings) \cite{wgc2024gold}.
We do not dwell on these comparisons, because a token ledger is not a
sovereign economy and its addresses are not households; the more
informative benchmark is the one the next subsection develops, which treats
the ledger as clearing infrastructure and asks how concentrated its routing
is.

\subsection{Market Structure: Routing Contestability}
\label{sec:hhi}
Rather than comparing a token ledger to the macro-wealth of a country, a more
logically rigorous approach is to evaluate it as a \textbf{financial
clearing and payment infrastructure}. This functional alignment requires a
shift from social wealth distribution metrics (Gini) to industrial market
concentration measures---specifically, the \textbf{Herfindahl--Hirschman
Index (HHI)} \cite{doj2010horizontal}. The relevant thresholds were revised in
the interval between that literature and this study, and we report against
both. The 2010
Horizontal Merger Guidelines \cite{doj2010horizontal} classified a market as
\textit{unconcentrated} below $1{,}500$, \textit{moderately concentrated}
between $1{,}500$ and $2{,}500$, and \textit{highly concentrated} above
$2{,}500$. The 2023 Merger Guidelines \cite{doj2023merger}, issued jointly by
the DOJ and FTC on 18 December 2023 and superseding both the 2010 horizontal
and the 2020 vertical guidelines, lower the structural presumption to a
post-merger HHI above $1{,}800$, restoring the threshold in force from 1982
to 2010, under which $1{,}000$--$1{,}800$ was the moderate band.

Our substantive conclusions do not turn on which set is applied, and we say
so explicitly because the revision postdates much of the literature we
benchmark against. Every address-level and entity-resolved figure we report
for either token---balance HHI of $135$ and $872$, $562$ and $962$ after
entity resolution, and weekly flow HHI of $421$ and $386$, $654$ and $667$
after entity resolution---falls below $1{,}000$ and is therefore
unconcentrated on both conventions. Two figures change band. LINK's peak
address-level week, at $1{,}147$, is unconcentrated under the 2010 thresholds
but moderately concentrated under those of 2023; and restricting UNI to
transfer-active addresses, which yields $2{,}336$, moves from moderately to
\emph{highly} concentrated. The entity-resolved LINK window at $1{,}509$ is
moderately concentrated under either convention and does not change band. Both
changes sharpen rather than soften the specification-sensitivity argument of
Section~\ref{sec:contrib-spec}.

\begin{figure*}[t]
  \centering
  \includegraphics[width=0.95\linewidth]{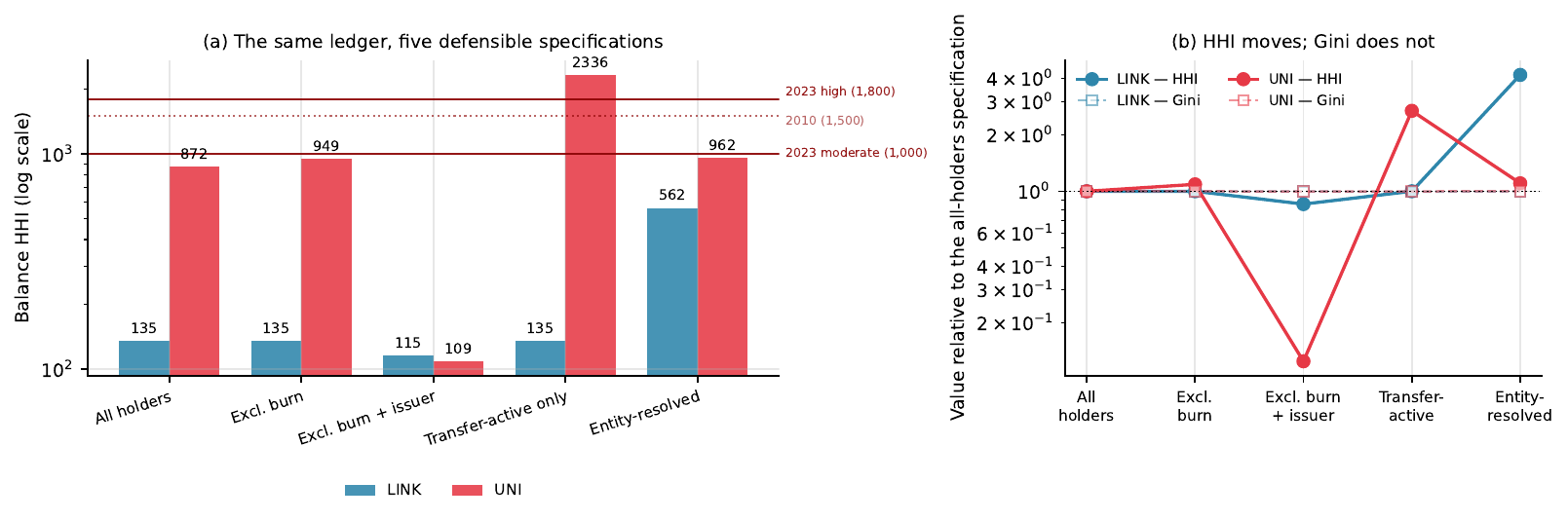}
  \caption{The same ledger under five defensible specifications. (a) Balance
  HHI on a logarithmic scale. Every bar is computed from the same address
  ledger; the specifications differ only in which holders are counted and
  whether an address or an entity is the unit of analysis. The range spans
  $115$--$562$ for LINK and $109$--$2{,}336$ for UNI, a factor of five and of
  twenty-one respectively. (b) The same specifications expressed relative to
  the all-holders baseline. The HHI traces move across the vertical axis
  while the Gini markers sit on the unit line throughout; under entity
  resolution the Gini does not move at all to four decimal places. The two
  statistics are not substitutes, and reporting one does not constrain the
  other.}
  \label{fig:specification}
\end{figure*}

\subsubsection{Computation Methodology}
Following standard antitrust practice, each unique Ethereum address is
treated as one market participant. Two HHI scores are reported, both on the
standard DOJ $0$--$10{,}000$ scale as $\sum_i (100 s_i)^2$.

\begin{itemize}
    \item \textbf{Balance HHI}: the share of address $i$ is
    $s_i = b_i / \sum_j b_j$, where $b_i$ is the token balance at the end of
    the window and the denominator sums over \emph{all} addresses with
    strictly positive balance ($885{,}011$ for LINK, $386{,}029$ for UNI).
    This is the same population used for the Gini coefficient and the
    top-$k$ shares reported alongside it, so that the four ownership
    statistics describe one and the same set of addresses.
    \item \textbf{Flow HHI}: within each weekly window $w$, the share of
    address $i$ is $s_{i,w} = f_{i,w} / \sum_j f_{j,w}$, where $f_{i,w}$ is
    the total outflow sent by $i$ during $w$, computed on the aggregated
    edge weights of Equation~\ref{eq:edgeweight}. Only sender-side flows are
    counted, to avoid double-counting each transfer. The reported
    $HHI_{\text{Flow}}$ is the mean of the $13$ weekly scores.
\end{itemize}

Two aspects of this construction are discretionary rather than dictated by
the data, and we make them explicit because both turn out to matter. First,
temporal aggregation: averaging thirteen weekly scores is not the same as
computing one score on the pooled 90-day flow, and the two differ by a factor
of $1.83$ for LINK ($421$ versus $230$) and $1.39$ for UNI ($386$ versus
$277$). We report the weekly mean because market structure is a property of
the period over which participants actually compete, and a single 90-day pool
averages away the entry and exit that concentration is meant to capture; the
pooled figure appears in \ref{app:sensitivity}. Second, and more
consequentially, standard antitrust practice treats each participant as one
economic actor, whereas an address is not an actor but an account. We return
to this in Section~\ref{sec:entity}.

\subsubsection{Empirical Results}
Our calculations yield a balance HHI of \textbf{135} for LINK and
\textbf{872} for UNI, and a mean weekly flow HHI of \textbf{421} for LINK and
\textbf{386} for UNI. All four fall below the DOJ unconcentrated threshold.
Fig.~\ref{fig:specification} places these figures among the alternative
specifications discussed in Section~\ref{sec:entity} and
\ref{app:sensitivity}.

\begin{figure}[H]
  \centering
  \includegraphics[width=0.98\linewidth]{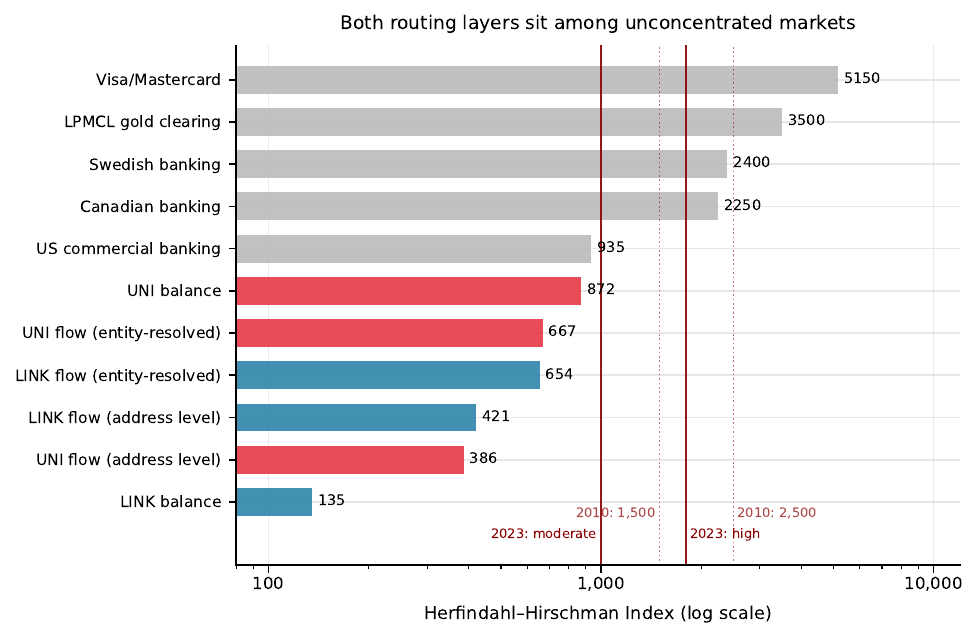}
  \caption{Herfindahl--Hirschman Index benchmarked against traditional
  clearing infrastructure, on a logarithmic scale. Our measurements are
  shown in colour: LINK balance ($135$), UNI balance ($872$), and weekly
  flow for both ledgers at address level ($421$, $386$) and after partial
  entity resolution ($654$, $667$). Reference points are U.S.\ commercial
  banking ($935$, the lower end of a $935$--$1{,}060$ range across recent
  years) \cite{ffiec2023bhc}, Canadian and Swedish banking ($2{,}250$ and
  $2{,}400$; indicative national figures, reported here for order of
  magnitude rather than as precise estimates), LPMCL gold clearing
  ($3{,}500$) \cite{lbma2023clearing}, and Visa/Mastercard ($5{,}150$)
  \cite{nilson2024cards}. Solid lines mark the 2023 Merger Guidelines
  thresholds \cite{doj2023merger}; dotted lines the 2010 Horizontal Merger
  Guidelines thresholds they replaced \cite{doj2010horizontal}. Under the
  2023 convention US commercial banking sits just below the moderate band,
  while the Canadian, Swedish, gold-clearing and card-clearing benchmarks are
  all highly concentrated. Routing concentration for both tokens sits
  inside the unconcentrated band despite extreme ownership inequality, and
  remains there after entity resolution, with the exceptions of LINK's entity-resolved window at $1{,}509$ and, under the 2023 thresholds, its peak address-level week at $1{,}147$.}
  \label{fig:hhi_comparison}
\end{figure}

\begin{itemize}
    \item \textbf{Asset holding (LINK $135$ vs.\ UNI $872$).} Despite Gini
    coefficients of $0.990$ and $0.998$, both balance distributions are
    classified as \textit{unconcentrated}; LINK sits below the U.S.\
    commercial banking deposit market ($HHI \approx 935$)
    \cite{ffiec2023bhc}. The top $10$ addresses hold $32.9\%$ of LINK and
    $51.7\%$ of UNI supply.

    We caution against reading the six-fold gap between the two tokens as an
    economic difference. Issuer-controlled supply accounts for $21.0\%$ of
    observed LINK (seven wallets labelled as non-circulating reserve) and
    $28.0\%$ of observed UNI (the governance Timelock and the token
    distributor), and the UNI burn address holds a further $10.8\%$.
    Excluding burn and issuer addresses brings the two ledgers to $115$ and
    $109$ respectively---a difference of $6\%$. The headline contrast is
    therefore attributable mainly to how much of each supply the protocol
    holds itself, not to concentration among market participants.

    \item \textbf{Traffic routing (LINK $421$ vs.\ UNI $386$).} Both ledgers
    are classified as unconcentrated in routing. Weekly scores range from $158$ to
    $1{,}147$ for LINK and $173$ to $593$ for UNI. Under the 2010 thresholds no
    single week on either ledger reaches the moderate band; under the 2023
    thresholds LINK's peak week does, at $1{,}147$ against a moderate floor of
    $1{,}000$, while every UNI week remains below it. The revision therefore
    adds a specification under which one address-level window changes band
    without any entity resolution at all.

    \item \textbf{The Gini--HHI divergence.} The coexistence of extreme Gini
    ($0.990$, $0.998$) with low flow HHI ($421$, $386$) is the central
    quantitative finding. The two statistics answer different questions: the
    HHI is a sum of squared shares and is dominated by the largest few
    participants, while the Gini describes the shape of the cumulative
    distribution across all of them. A ledger can be simultaneously very
    unequal in the long tail and unconcentrated at the head. Traditional
    financial rails display the opposite pattern---card clearing
    (Visa/Mastercard, $HHI = 5{,}150$) \cite{nilson2024cards} and LPMCL gold
    clearing ($HHI \approx 3{,}500$) \cite{lbma2023clearing} are dominated by
    two to four entities commanding $25$--$90\%$ of their markets.
    Fig.~\ref{fig:hhi_comparison} places our measurements on a common scale
    with these benchmarks.
\end{itemize}

\subsubsection{Entity Resolution and the Limits of Address-Level Accounting}
\label{sec:entity}
Treating one address as one participant understates concentration whenever a
single operator controls several addresses, which on a public ledger is both
cheap and common. To bound this we resolve addresses to entities using the
archived label snapshot, merging addresses whose labels share an
operator---all wallets tagged \texttt{Binance}, all contracts tagged
\texttt{Uniswap}---while leaving generic tags such as \texttt{MEV Bot}
unmerged, since two bots are not one firm.

The effect is large and asymmetric across statistics. For LINK the balance
HHI rises from $135$ to $562$ ($4.16\times$), driven almost entirely by seven
equal-sized reserve wallets collapsing into one participant, while the
balance Gini is unchanged to four decimal places ($0.9898$ in both cases).
The same pattern holds for UNI ($872 \rightarrow 962$; Gini $0.9976$
unchanged). This is a general property rather than a quirk of these data:
because the HHI squares shares, merging the head is amplified
quadratically, whereas the Gini integrates over a distribution of nearly nine
hundred thousand addresses in which a handful of merges is invisible. Any HHI
computed on address-level blockchain data should be read as a range rather
than a point estimate.

For flow, entity resolution raises the mean weekly HHI from $421$ to $654$
for LINK ($1.55\times$) and from $386$ to $667$ for UNI ($1.73\times$). One LINK window reaches $1{,}509$---marginally above the 2010 threshold of
$1{,}500$, and comfortably inside the 2023 moderate band; the
remaining twelve windows, and all thirteen UNI windows, stay below it. The
largest single entity is the same on both ledgers: addresses tagged
\texttt{Binance} account for $18.4\%$ of mean weekly outflow on LINK across
five addresses and $19.8\%$ on UNI across five, contributing $337$ and $393$
HHI points---roughly half of the entity-level index on each ledger.

Two caveats bound the interpretation. First, our label snapshot covers
$33.5\%$ (LINK) and $31.5\%$ (UNI) of mean weekly outflow; unlabelled
addresses remain separate participants, so the entity-level figures are a
\emph{lower bound} on true concentration, not an adversarial upper bound.
Second, exchange addresses hold and move customer assets rather than their
own, so merging them measures custodial rather than beneficial
concentration. We therefore state the conclusion in the form the evidence
supports: under address-level accounting and the 2010 thresholds both routing layers
are unconcentrated in every window, while the 2023 thresholds place LINK's
peak week in the moderate band; under partial entity resolution LINK reaches
the moderate band in one window of thirteen; and the figure under complete
entity resolution is higher still. What does not change under any
specification is the distance between the ownership and routing layers.

\subsection{Structural Role Decomposition and the Wealth-Power Divergence}
The aggregate concentration metrics above describe the \textit{degree} of inequality, but they say little about \textit{who} occupies which position and why. The GNN clustering decomposes the network into eight functional roles that, taken together, paint a more granular picture of how value flows through the ledger.

\subsubsection{The Two-Layer Clearing Core}
The densest stratum of the network is a single small cluster. On LINK it comprises $169$ nodes with a median of $80$ counterparties and a core rate of $10.1\%$; on UNI, $154$ nodes, median $78$ counterparties, core rate $8.4\%$. It is populated by major DEX liquidity pools, CEX hot wallets and router contracts that settle the bulk of high-value transfers. A second, broader cluster ($1{,}492$ nodes on LINK, median $7$ counterparties, core rate $3.3\%$) serves as a wide-distribution buffer, fanning transactions out from the dense core to the broader user base. Together these form a ``hub-and-spoke'' topology functionally analogous to the correspondent-banking tiering observed in traditional interbank payment systems \cite{kindleberger2011manias}. The near-identical dimensions of the infrastructure cluster on two functionally unrelated ledgers is itself notable: roughly $1.6\%$ of the addresses in each network carry its clearing function.

\subsubsection{Wealth and Structural Power Are Orthogonal}
\label{sec:wealth-power}
The relationship between what an address holds and where it sits is best stated at the level of individual addresses, where it can be measured directly. Across addresses with a positive balance, the Spearman correlation between balance and counterparty count is $+0.081$ on LINK and $+0.032$ on UNI; between balance and betweenness it is $-0.028$ and $-0.051$. All four are indistinguishable from zero on samples of thousands of nodes. Wealth and structural power are not inversely related on these ledgers, nor positively---they are close to \emph{orthogonal}.

This orthogonality is the mechanism behind the paradox this paper documents. A Gini coefficient of $0.99$ constrains how token holdings are distributed; it says nothing whatever about who routes them, precisely because the two quantities are uncorrelated. Extreme ownership inequality and a contestable clearing layer are not in tension; they are independent facts about independent dimensions. In programmable ledger economies, holding capital does not confer operational control---a divergence qualitatively sharper than in equity markets, where large shareholders typically exert governance influence proportional to their stake.

\subsubsection{Zero-Balance Intermediaries}
Within the periphery the criterion isolates $30$ addresses on LINK and $18$
on UNI that hold no meaningful balance yet rank in the top percentile of
structural-core score. Counting each address once, manual verification resolves $14$ of the $36$
distinct brokers---$39\%$---to non-custodial routing contracts (counting per
ledger, $20$ of $48$)---DEX aggregators, wallet swap interfaces,
Uniswap's Universal Router and X solver, cross-chain routers---that move
liquidity without ever taking custody. They operate on fees forwarded
immediately to off-chain treasuries, which is why they are invisible to any
ranking based on holdings. The remainder show the same behavioural signature but carry no identity in
the archived label snapshot; a few of them, such as the MEV bots and the
phishing-flagged address, do carry labels on the live Etherscan site. Because wealth and
structural position are statistically independent across the address
population (Section~\ref{sec:wealth-power}), a framework attending to only one
of the two will mischaracterise the distribution of power on these ledgers.

\subsection{A Case in Which an Implementation Detail Reversed a Finding}
\label{sec:edgeweight-case}
Sensitivity analyses usually argue from hypothetical specifications. This
one does not. In the course of preparing this paper we discovered that an
earlier version of the analysis had constructed the transfer graph in a way
that discarded most of its weight, and that correcting it removed one of the
paper's stated findings. Because such episodes are rarely reported, and
because this one bears directly on the argument of
Section~\ref{sec:contrib-spec}, we set it out in full.

\paragraph{The detail}
Repeated transfers between the same ordered pair of addresses must be summed
into a single weighted edge (Equation~\ref{eq:edgeweight}). Standard graph
libraries, when handed a transfer list, build a simple directed graph in
which a repeated edge \emph{overwrites} its predecessor's attribute rather
than accumulating it. No error is raised; the node set, the edge set and the
degree sequence are all unchanged. Only the weights differ.

\paragraph{The magnitude}
Repeat interaction is the dominant mode of activity on both ledgers. Across
the full ledger, $36.7\%$ of LINK records and $83.9\%$ of UNI records are
repeat edges, rising to $90.9\%$ and $91.7\%$ within the activity-ranked core
subgraphs. An unaggregated construction therefore discards $85.6\%$ and
$85.3\%$ of transferred volume in those subgraphs. PageRank computed the two ways correlates at only $0.83$ on LINK and $0.70$
on UNI, and the top fifty addresses by PageRank overlap in $31$ and $33$ of
$50$ positions.

\paragraph{The reversal}
Under the unaggregated construction, the infrastructure cluster on LINK
recorded a median balance of $0.33$ while the highest-balance cluster
recorded a core rate of $0.3\%$. That pattern supported a finding we had
written into the manuscript: on an oracle token, wealth and clearing
function are borne by \emph{disjoint} clusters, in contrast to a governance
token where they coincide. Under the corrected construction the same
clustering procedure returns an infrastructure cluster with a median balance
of $3{,}934$ on LINK and $3{,}405$ on UNI, and on both ledgers that cluster
is simultaneously the highest in core rate and in median balance. The
cross-token contrast does not survive; we withdrew it.

The claim that replaced it is stronger and is stated at the level where it
is actually measurable. Across individual addresses with a positive balance,
balance and structural position are close to statistically independent
(Section~\ref{sec:wealth-power}), which is what permits extreme ownership
inequality to coexist with an unconcentrated clearing layer. The
cluster-level pattern that the faulty construction had produced was an
artefact; the address-level independence is not.

\paragraph{Why this matters beyond our own correction}
Three features of this episode generalise. First, the faulty construction
produced results that were internally consistent and superficially
plausible---a two-layer clearing core, a passive whale stratum---which is
precisely why it survived several rounds of review. Second, it was invisible
to every check that did not involve weights: node counts, anchor counts and
cluster sizes all matched. Third, the resulting error was not uniform across
statistics. Flow HHI, computed by direct aggregation rather than through the
graph, was unaffected and matched to within $0.5$ points across all thirteen
weekly windows; the embedding-derived quantities were substantially wrong.
A robustness check confined to the statistics that happened to be computed
correctly would have found nothing.

We archive the superseded results alongside the corrected ones so that the
comparison can be reproduced rather than taken on trust.

\subsection{The Rule Layer: Who Can Revoke Contestability}
\label{sec:governance}
The two scales analysed so far answer two questions---who owns the supply,
and who moves it. A natural objection raises a third that neither
can address: on a token whose holders vote in proportion to their balance,
what is the standing of a contestable routing layer if a small cohort retains
the authority to change the rules under which that routing occurs? The
question is well founded. The HHI measures who is doing business; it does
not measure who can amend the terms of business. We therefore add a rule
layer to the analysis, using an established instrument rather than a new one.

Because governance is a threshold game---a proposal either reaches quorum or
it does not---the appropriate statistic is not a dispersion index but the
Nakamoto coefficient \cite{grajales2022measuring}, the minimum number of
participants whose combined holdings reach a decision threshold. Uniswap
governance sets a proposal threshold of $2.5\times10^6$ UNI ($0.25\%$ of
supply) and a quorum of $4\times10^7$ ($4\%$).

\paragraph{The coefficient is as specification-dependent as the HHI}
Counting all addresses, a single address holds enough UNI to meet quorum
unaided. That address is the governance Timelock itself, and the second
largest is the burn address; neither can vote. Excluding addresses that
cannot cast a vote raises the coefficient to two, and excluding custodial
exchange addresses---which hold customer assets and do not, as a rule, vote
them---leaves it at two while reducing the number of addresses individually
above the proposal threshold from $55$ to $47$ and their combined share from
$73.8\%$ to $29.1\%$. The unadjusted $73.8\%$ figure, which an earlier draft
reported, is $38.7$ percentage points inflated by the Timelock and the burn
address; the remaining $6.0$ points are removed with the token distributor
and the custodial exchange wallets. The same discipline the HHI requires is required here.

One direction of error in this figure is worth stating, because it runs
against us rather than for us. A balance-based Nakamoto coefficient is a
\emph{conservative upper bound} on governance dispersion. Under Uniswap's
Compound-derived architecture, voting weight is not conferred by holding but
by explicit delegation through \texttt{delegate()}; un-delegated tokens carry
zero weight at a proposal's checkpoint block. Because a substantial share of
circulating supply sits in cold wallets that never delegate, the pool of
addresses with live voting weight is strictly smaller than the set we count.
Effective control is therefore more concentrated than $N=2$ indicates, not
less.

\paragraph{The rule layer and the routing layer are largely disjoint}
Fifty-two addresses hold enough UNI to open a proposal and are capable of
voting---that is, after removing the governance Timelock, the token
distributor and the burn address from the $55$ that clear the threshold on
balance alone. Thirty-seven of those fifty-two---$71\%$---sent and received
nothing at all during the 90-day window. To test whether those who make the rules also
operate the network, we compare the eligible set against the routing layer
defined by weekly outflow, which imposes no balance condition of any kind.
Among the ten largest senders by mean weekly outflow, two are
proposal-eligible; among the top fifty, five; among the top hundred, seven.
The median balance of the fifty largest senders is $45{,}504$ UNI, fifty-five
times below the proposal threshold. The overlap that does exist consists
almost entirely of exchange addresses, which move customer assets and hold
them on customers' behalf.

We stress the construction of this test because a natural alternative is
circular. Comparing proposal-eligible addresses against the hidden brokers
of Section~\ref{sec:c3} would yield an overlap of exactly zero, but only
because a hidden broker is \emph{defined} as holding at most a median
balance while proposal eligibility \emph{requires} $2.5\times10^6$ tokens.
That comparison cannot fail, and therefore says nothing. The outflow-based
test can fail and does not.

\paragraph{How much of the routing layer is actually revocable}
The objection has a quantitative form: if Uniswap governance can
rewrite the rules the routing layer follows, its contestability is
conditional. We bound the exposure on the dimension we observe. Of mean
weekly UNI outflow, $0.71\%$ originates from contracts the governance
Timelock directly controls, and a further $0.14\%$ from other
Uniswap-deployed contracts, most of which are immutable and cannot be
amended by governance in any case. A remaining $68.5\%$ flows from addresses
our label snapshot does not identify, which bounds the upper end of the range
rather than establishing it.

That residual is not, however, a blank cheque for the upper bound. Governance
can amend only the contracts it deploys and retains upgrade authority over;
it has no mechanism whatsoever for altering the behaviour of an externally
owned account, and the overwhelming majority of unlabelled addresses on a
transfer ledger are externally owned accounts rather than upgradeable
contracts. The share of the residual that could in principle fall under
governance control is therefore bounded by the share of it that is an
upgradeable protocol contract, which is a small fraction of the whole. The
true figure lies much closer to the lower bound than the arithmetic width of
the interval suggests.

This measurement has a boundary that must be stated plainly. It covers UNI
\emph{token transfers}, not swap volume in Uniswap pools. A governance
decision to activate the protocol fee switch would affect trading that
settles in those pools regardless of which router carried it, and that
channel lies outside our data. What we can say is narrower and still
informative: on the dimension of UNI token routing, the share directly under
governance control is under one percent.

\paragraph{Contestability is conditional, not structural}
Taken together these results qualify the paper's central finding rather than
overturning it. The routing layer is genuinely contestable: many mutually
competing intermediaries, low measured concentration, weekly turnover in the
leading positions. But on a token-weighted governance ledger that
contestability sits inside a rule layer that a small, largely inactive cohort
can amend. The two layers are almost disjoint in membership, which means
neither can be inferred from the other---and it means that a
concentration statistic computed on either one alone will mischaracterise
the system.

The result compresses to a single sentence. UNI's routing layer records a
flow HHI of $386$ and is unconcentrated by every antitrust criterion we can
apply to it; rewriting the rules that layer operates under requires the
coordination of two voting-capable addresses. Contestability in the routing
layer is held at the pleasure of a rule layer with a Nakamoto coefficient of
two.

Three limits apply to the governance figures. Holding is not voting: UNI
requires explicit delegation, so our counts measure an upper bound on
potential voting power. Uniswap's own governance documentation reports that
many large delegates participate well below half the time and some not at
all, which implies that effective decision power is more concentrated than
our upper bound, not less. Custodial addresses do not vote their holdings and
are reported separately for that reason. Finally, with two tokens and only
one of them token-governed, we present this as a description of a single
protocol's rule layer, not as a detector that classifies protocols. LINK has
no token-weighted governance---oracle node operators are admitted by
Chainlink Labs---so the comparison is itself informative: of two ledgers
whose routing layers are equally contestable, only one has a rule layer
capable of revoking that contestability.

\subsection{Which Layer Is the Claim About?}
``Decentralized'' is a predicate that requires a layer. At the base
protocol---consensus rules, permissionless deployment, custody---our evidence
supports the usual verdict on both ledgers: no participant commands a
dominant share, and neither the largest holders nor the busiest routers can
rewrite consensus, freeze balances, or bar a competing contract from being
deployed. That verdict does not extend upward. A protocol may be credibly
neutral at its base while running an application whose parameters a small
cohort can amend by vote, and UNI is such a case
(Section~\ref{sec:governance}). Rule neutrality guarantees that nobody can
seize your tokens; it guarantees nothing about who sets the fee on the pool
you trade in.

Two further exposures follow from the measurements above and are invisible to
a value-weighted index. Governance is plutocratic where voting is
token-weighted, since a balance Gini of $0.998$ concentrates the capacity to
propose in a handful of addresses. And the routing layer, though
unconcentrated by market share, is functionally dependent on a small set of
shared contracts: a bug in one of the twelve brokers common to both ledgers
would disrupt clearing on two tokens at once, and six of those twelve carry
no public label under which a supervisor could find them.

\section{Limitations}
\label{sec:limitations}

\textbf{Scope of the comparison.} We analyse two tokens. All three
contributions are replicated on both---the specification analysis, the
hidden-broker criterion with its matched control, and the three-layer
decomposition, the last of which is necessarily confined to UNI at the rule
layer---but two ledgers cannot
establish universality across an asset class of thousands. Holding the
observation window fixed removes market-regime effects as a confound, and
the convergence of two functionally dissimilar assets on the same routing
verdict is stronger evidence than a single case would provide; it
nonetheless remains a comparison of two. The UNI
transfer graph is also an order of magnitude smaller than LINK's
($38{,}667$ versus $642{,}126$ active addresses), so the head-to-head
structural comparison is conducted at the matched $10^4$ scale rather than
at full scale.

\textbf{Temporal coverage.} The 90-day window captures rich dynamics,
including a coordinated phishing event, but cannot speak to concentration
dynamics across full market cycles. Whether the routing contestability we
document survives a prolonged bear market remains open.

\textbf{Label coverage and label drift.} Newer infrastructure is
systematically unlabelled in the public tag compilation we use. This works
against us rather than for us---several routers recovered by the GNN
postdate the snapshot---but it means the labelled positive set understates
true infrastructure prevalence. A second and less obvious issue is that the
compilation is updated continuously and silently: it held $29{,}945$ entries
on 5 August 2026 and $173$ fewer four days later. Since the \texttt{is\_core}
anchor set is derived from it, an analysis run on two different days can
yield different AUCs with no change to the code or the data. We fix a single
snapshot and archive it (\ref{app:repro}); studies that query
such a database live should expect their supervised results to drift.

\textbf{Balance is not time-resolved.} Balances are cumulative historical
positions rather than per-window snapshots. Temporal analyses therefore use
within-window flow rather than balance, and the two scales should not be
read as measuring the same quantity at the same instant.

\textbf{Entity resolution is partial, so concentration is a lower bound.}
Our label snapshot covers about a third of weekly outflow on each ledger.
Addresses it does not identify remain separate participants in the
entity-level HHI, which therefore understates true concentration. We label
this a lower bound throughout and avoid the phrase ``upper bound'' even
where an adversarial reading might be tempting.

\textbf{Concentration metrics are protocol-agnostic.} Neither the Gini
coefficient nor the HHI distinguishes a protocol treasury contract from a
private whale. This is not a hypothetical concern: issuer-controlled supply
is $21.0\%$ of observed LINK and $28.0\%$ of observed UNI, and removing it
along with the burn address brings the two ledgers' balance HHI from $135$
and $872$ to $115$ and $109$. We report the unadjusted figures as the main
specification, because deciding which contracts are ``protocol-controlled''
is itself a judgement, and give the adjusted grid in
\ref{app:sensitivity}. Readers comparing our ownership figures with
other studies should check which convention those studies adopt; the choice
moves the number by a factor of eight on UNI.

\FloatBarrier

\textbf{Points developed in the main text.} Four further limitations are
discussed where they arise and we only index them here: sampled betweenness
degenerates on sparse graphs (Section~\ref{sec:c1}); the GNN's AUC
advantage over hand-crafted features does not survive scaling, and the
broker criterion owes its performance to counterparty count rather than to
the representation (Sections~\ref{sec:c1} and~\ref{sec:control});
broker membership is stable in aggregate but not per node
(Section~\ref{sec:brokerstability}); and manual verification must use the
token-transfer view rather than the transaction view, which on routing
contracts can differ by three orders of magnitude (\ref{app:repro}).

\section{Conclusion \& Future Work}
\label{sec:conclusion}

\subsection{Conclusion}
This paper asked what it takes to answer the question a great deal of
empirical work on blockchains takes for granted: is this ledger
concentrated? Across two functionally dissimilar ERC-20 ledgers---the
Chainlink (LINK) oracle token and the Uniswap (UNI) governance token,
observed over an identical 90-day window---we find that the answer is
determined less by the ledgers than by four measurement choices that no data
can settle, and that the range those choices span (a factor of twenty-one on
one of our two tokens, Fig.~\ref{fig:specification}) is wider than most of
the differences such studies set out to detect.

On the measurement side, the central finding is a sharp Gini--HHI divergence: both ledgers exhibit extreme wealth inequality ($Gini = 0.990$ and $0.998$), yet when evaluated as industrial clearing infrastructure their balance and flow HHI ($135$ and $421$ for LINK, $872$ and $386$ for UNI) fall in the DOJ \textit{unconcentrated} zone---below traditional payment incumbents such as Visa/Mastercard ($HHI = 5{,}150$) and physical gold clearing ($HHI \approx 3{,}500$). This divergence demonstrates that token wealth polarization does not imply monopolistic market power: the large number of active participants disperses market share below antitrust thresholds even as the cumulative distribution is severely skewed, and the two quantities are close to statistically independent across addresses. The divergence is bounded, not unconditional---partial entity resolution raises the flow figures to $654$ and $667$, and on a token-governed ledger the routing layer's contestability is itself revocable by a rule layer with which it barely overlaps. Each of these numbers is a specification as much as a measurement, which is why we report them as ranges and state the choices that produce them.

On the structural side, the structural analysis reinforces this paradox from a micro-structural perspective. The zero-balance hidden brokers identified by the criterion---$30$ on LINK and $18$ on UNI, of which $14$ of the $36$ distinct addresses resolve to named routing protocols---command disproportionate topological influence by bridging Burt's structural holes, while balance and structural position are statistically uncorrelated across the address population as a whole. Power in the network is thus neither proportional to wealth (as in equity markets) nor proportional to market share (as in classical oligopolies), but driven by structural positioning.

Taken together, these findings support a qualified conclusion. Clearing
competition on both ledgers is structurally open: no single participant
approaches a dominant share of weekly routing, and the leading positions turn
over week to week. But that openness is a property of one layer, and on a
token-weighted ledger it sits inside a second layer that can amend it. The
fifty-two UNI addresses able to open a governance proposal and the addresses
that actually route the token are almost disjoint sets, and seventy-one
percent of the former transacted not at all during the window. Contestability
in the routing layer and authority in the rule layer are held by different
people.

Whether that configuration counts as ``decentralization'' is not a question
a concentration statistic can settle, and we do not propose a new index to
settle it. What the evidence does establish is that the question cannot be
answered from any single number: the same ledger is unconcentrated by flow
HHI, extremely unequal by balance Gini, and governed by a cohort small enough
to count. Each is correct, each is incomplete, and a framework that reports
only one of them will mischaracterise the system---which is what motivated
the layered measurement we have set out here.

\paragraph{The research questions in brief}
RQ1: ownership is extreme and routing unconcentrated on both ledgers, and the
raw gap in balance HHI between them is a matter of issuer holdings. RQ2:
hidden brokers exist, are shared across ledgers, and half of the shared set
carries no public identity; the criterion, not the representation, does the
discriminative work (Section~\ref{sec:control}). RQ3: both ledgers resolve
one small infrastructure cluster above a retail periphery. RQ4: the phishing
airdrop saturated a dispersion measure without moving market structure.
RQ5: every verdict depends on specification, most strongly on entity
resolution, and on the governance ledger routing contestability is revocable
by a rule layer with a Nakamoto coefficient of two.
\label{sec:rq}

\subsection{Future Work}
\label{sec:future}
Four directions follow from the limitations above. First,
\textbf{propagation at full ledger scale}: our analysis is conducted on
activity-ranked subgraphs because both the embedding and the centrality
baselines it is compared against become costly on the complete transfer
graph, and approximate propagation methods~\cite{wang2021agp} are the
natural route to removing that restriction. Doing so would also allow the
sampled betweenness baseline to be replaced with something that does not
degenerate on sparse graphs, which would sharpen rather than soften the
comparison we draw. Second, \textbf{reverse-engineering solver and
aggregator routing} via on-chain call-trace analysis, to estimate solver
risk premiums and bidding behaviour for the intermediaries this paper can
only identify structurally. Third, \textbf{multi-chain and cross-layer flow
analysis}, extending the framework to measure interoperability concentration
across bridges and Layer-2 rollups---particularly relevant given that
several of the shared brokers we recover carry cross-chain extensions.
Fourth, \textbf{dynamic temporal architectures} (T-GCN, EvolveGCN) capable
of detecting a routing anomaly as it forms rather than in retrospect, as our
weekly windowing necessarily does.

\section*{CRediT Authorship Contribution Statement}
\textbf{Jintao Liu:} Conceptualization, Methodology, Software, Formal
analysis, Data curation, Investigation, Visualization, Writing --- original
draft, Writing --- review \& editing. 
\textbf{Zhimo Ji:} Conceptualization, Methodology, Formal analysis, 
Investigation, Writing --- original draft, Writing --- review \& editing. 
\textbf{Xuzhe Lin:} Data curation, Validation, Writing --- review \& editing. 
Jintao Liu and Zhimo Ji contributed equally to this work.

\section*{Declaration of Competing Interest}
The authors declare that they have no known competing financial interests
or personal relationships that could have appeared to influence the work
reported in this paper. Neither author holds a financial position in the
Chainlink (LINK) or Uniswap (UNI) tokens analysed here, nor any affiliation
with the protocols, exchanges, or routing services identified in the
analysis.

\section*{Data Availability}
All primary data are drawn from the public Ethereum ledger and are
reproducible by anyone. Transfer records and balances were extracted from
the \url{bigquery-public-data.crypto_ethereum.token_transfers} dataset
via Google BigQuery; the exact extraction queries, the observation window
(2026-03-26 to 2026-06-24), and the two token contract addresses are
specified in Section~\ref{sec:data}. Entity labels are taken from a
single archived snapshot of a public Etherscan label compilation (retrieved
26 August 2026, MD5 \texttt{16792aac5afd}), which is deposited with the code
rather than queried live, for the reason given in
\ref{app:repro}. Derived artefacts---node features, the learned
embeddings underlying every reported ensemble, cluster assignments, and the
identified broker sets---together with the analysis code are available at
\url{https://github.com/Liu0916-star/GNN-Wallet-Clustering}. A preprint of
this manuscript is deposited on arXiv under a CC BY-NC-ND license.

\section*{Funding}
This research received no specific grant from any funding agency in the
public, commercial, or not-for-profit sectors.

\FloatBarrier
\appendix
\setcounter{table}{0}
\setcounter{figure}{0}
\renewcommand{\thetable}{A.\arabic{table}}
\renewcommand{\thefigure}{A.\arabic{figure}}
\section{Sensitivity of the Concentration Measures}
\label{app:sensitivity}

Computing an HHI over an address ledger requires four discretionary
choices. Table~\ref{tab:sensitivity} reports how each affects the result for
both tokens.

\begin{table}[H]
\centering\footnotesize
\caption{Sensitivity of the balance and flow HHI to four discretionary
measurement choices. Values in bold are the main-text specification. Two
specifications place a token above the 2010 unconcentrated threshold of
$1{,}500$: restricting UNI to transfer-active addresses ($2{,}336$), and
entity resolution on LINK, which crosses it in that ledger's second weekly
window ($1{,}509$) though not in the mean of the thirteen. Under the 2023
thresholds the first is highly rather than moderately concentrated, and
LINK's peak address-level window ($1{,}147$) also enters the moderate band.}
\label{tab:sensitivity}
\setlength{\tabcolsep}{4pt}
\begin{tabular}{@{}llrr@{}}
\toprule
Choice & Specification & LINK & UNI \\
\midrule
\multirow{2}{*}{Population (balance)}
  & All positive balances      & \textbf{135} & \textbf{872} \\
  & Transfer-active only       & 135          & $2{,}336$ \\
\addlinespace[2pt]
\multirow{3}{*}{Address type (balance)}
  & All holders                & \textbf{135} & \textbf{872} \\
  & Excluding burn             & 135          & 949 \\
  & Excluding burn and issuer  & 115          & 109 \\
\addlinespace[2pt]
\multirow{2}{*}{Temporal aggregation (flow)}
  & Weekly, then averaged      & \textbf{421} & \textbf{386} \\
  & Pooled over full window    & 230          & 277 \\
\addlinespace[2pt]
\multirow{2}{*}{Flow direction}
  & Sender-side outflow        & \textbf{421} & \textbf{386} \\
  & Receiver-side inflow       & 418          & 386 \\
\addlinespace[2pt]
\multirow{2}{*}{Entity resolution}
  & One address, one participant & \textbf{421} & \textbf{386} \\
  & Labelled addresses merged    & 654          & 667 \\
\bottomrule
\end{tabular}
\end{table}

Five observations follow: one for each of the four choices, plus a separate
treatment of the burn address, which falls under address type but behaves in
a way that is easy to misread.

First, flow direction is immaterial: outflow and inflow specifications agree
to within $1\%$ on both tokens, so the result does not depend on which side
of the transfer is credited.

Second, temporal aggregation matters and in the conservative direction.
Pooling the full window \emph{lowers} measured concentration by a factor of
$1.83$ on LINK and $1.39$ on UNI, so the weekly-averaged figure we report is
the more concentrated of the two.

Third, population scope matters only for UNI, where restricting to
transfer-active addresses raises the balance HHI from $872$ to $2{,}336$.
Only $4\%$ of UNI holders transacted during the window, so the active subset
is dominated by treasury and routing contracts. For LINK, where $64\%$ of
holders were active, the choice is immaterial to one decimal place.

Fourth, excluding the burn address \emph{raises} UNI's balance HHI, from
$872$ to $949$. This is arithmetic rather than anomaly. The burn address
holds $10.76\%$ of observed supply, contributing $10.76^2 = 115.8$ points;
removing it shrinks the denominator by the same factor, inflating every
remaining share by $1/0.892$ and every squared share by $1.256$. The
residual $872 - 115.8 = 756.2$ therefore rescales to $949.4$, which is what
we observe. LINK's burn address holds $110$ tokens and its measures are
unchanged to four decimal places. Excluding issuer-controlled supply as
well brings the two ledgers to $115$ and $109$: a seven-fold gap in the
headline figures collapses to a difference of $6\%$ once protocol holdings
are set aside.

Fifth, and most consequentially for interpretation, the HHI is highly
sensitive to entity resolution while the Gini and top-$k$ shares are not.
Chainlink distributes its non-circulating reserve across seven wallets of
$3\times10^7$ LINK each; treating them as one participant raises the LINK
balance HHI from $135$ to $562$, because the squared-share functional form
amplifies head aggregation quadratically. Over the same specifications LINK's
balance Gini moves only from $0.9898$ to $0.9871$---a change of $0.0027$,
against a factor of four in the HHI---and its top-$1\%$ share from $93.2\%$ to
$91.4\%$. Under entity resolution specifically, the Gini does not move at all
to four decimal places while the HHI more than quadruples. Because the number of participants is a
free parameter under address-level accounting, an HHI computed on blockchain
data should be read as a bounded range rather than a point estimate. Across
the full range considered here, $115 \le HHI_{\text{Balance}} \le 562$ for
LINK and $109 \le HHI_{\text{Balance}} \le 2{,}336$ for UNI, and
$230 \le HHI_{\text{Flow}} \le 654$ and $277 \le HHI_{\text{Flow}} \le 667$
respectively. Every mean flow specification remains inside the
unconcentrated band on both conventions. On the balance side the exception is
UNI restricted to transfer-active addresses ($2{,}336$): moderately
concentrated under the 2010 thresholds, highly concentrated under those of
2023. Among individual weekly windows, LINK's second window under entity
resolution reaches $1{,}509$, and its peak address-level window reaches
$1{,}147$, which the 2023 thresholds also place in the moderate band.

The Gini coefficients reported throughout use the standard discrete
estimator over addresses with strictly positive balance; two independent
implementations used during the analysis agree to six decimal places.

\FloatBarrier
\section{Supporting Evidence for the Instrument}
\label{app:instrument}

This appendix collects the scale and ablation evidence supporting
Section~\ref{sec:c1}. It is placed here rather than in the main text because
its role is to establish the limits of the operationalisation we adopt, not
to advance a claim of its own.

\paragraph{Predicting top-balance nodes}
We next predict whether a node belongs to the top 5\% by balance. Because
balance is removed from the GNN input, this remains a leakage-free test of
whether structural position alone encodes economic weight.
Table~\ref{tab:c1_balance} shows that the GNN dominates at every scale,
reaching AUC $0.928$ at the largest. Unlike the \texttt{is\_core} target,
here the GNN's advantage over the structural features is large
($+0.147$, $+0.058$, $+0.045$) and survives seed variation at every scale.
We note one inversion worth recording: at $10^4$ the two-dimensional
centrality baseline ($0.770$) marginally outperforms the seven-dimensional
structural set ($0.767$), so a richer hand-crafted feature vector is not
uniformly better for this target.

\begin{table}[H]
\centering\small
\caption{five-fold cross-validated AUC for predicting
top-5\% balance nodes. Balance is excluded from the GNN input. GNN figures
are means over random initialisations with the seed-level standard
deviation.}
\label{tab:c1_balance}
\begin{tabular}{lccc}
\toprule
Scale & Centrality & Full struct. & \textbf{GNN} \\
\midrule
$10^4$          & $0.770$ & $0.767$ & $\mathbf{0.914 \pm 0.017}$ \\
$5\times10^4$   & $0.792$ & $0.812$ & $\mathbf{0.870 \pm 0.017}$ \\
$6.4\times10^5$ & $0.845$ & $0.883$ & $\mathbf{0.928 \pm 0.013}$ \\
\bottomrule
\end{tabular}
\end{table}

\begin{figure}[H]
  \centering
  \includegraphics[width=\linewidth]{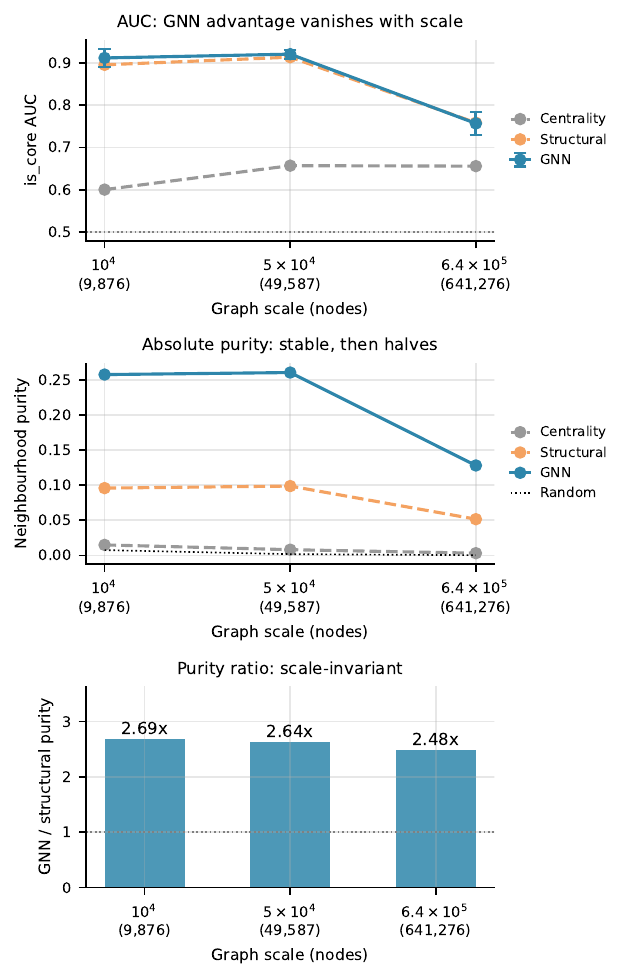}
  \caption{The instrument across three graph scales on LINK. (a) AUC for the
  external \texttt{is\_core} label: classical centrality stays near chance
  while the GNN's margin over dimension-matched structural features narrows
  from $+0.016$ to $-0.002$ and never exceeds one seed-level standard
  deviation. (b) Absolute neighbourhood purity, plotted against the random
  baseline (dotted), which falls by a factor of $30$ across the three scales;
  centrality's purity \emph{declines} even as its enrichment multiple rises.
  (c) The GNN-to-structural purity ratio, which is invariant to that baseline
  and holds between $2.48$ and $2.69$ across a $65$-fold range in node
  count.}
  \label{fig:c1_iscore}
\end{figure}

\begin{figure}[H]
  \centering
  \includegraphics[width=0.98\linewidth]{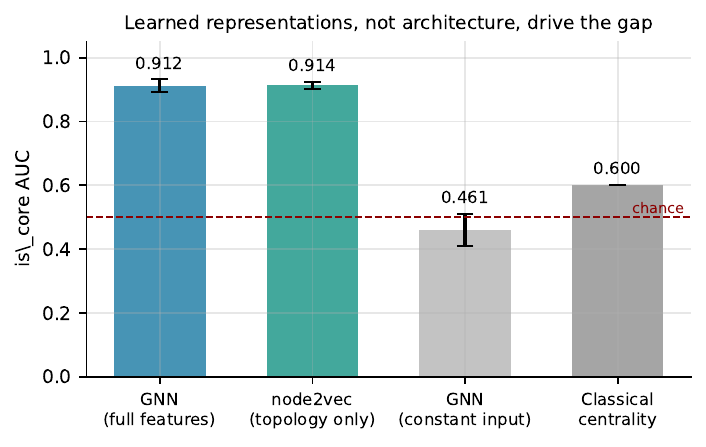}
  \caption{Ablation on the $10^4$ LINK graph. The GNN and node2vec are
  statistically indistinguishable on this target, and both far exceed
  classical centrality; the dashed line marks chance. The constant-input
  configuration falls below chance because mean aggregation over
  uninformative features returns near-identical vectors for all nodes---a
  degeneracy of the architecture under that input, not evidence that
  topology is uninformative, as the node2vec bar shows. Error bars are one
  standard deviation across random initialisations.}
  \label{fig:ablation}
\end{figure}

\section{The Week-7 Phishing Airdrop}
\label{app:phishing}
The thirty-eight contracts responsible for the anomalous LINK window are
funded from thirteen upstream addresses, and $94\%$ of their activity falls on
two consecutive days (9--10 May 2026). Each is a verified contract exposing
an \texttt{Airdrop} method and presenting itself as \texttt{LinkGiveaway} or
\texttt{LinkStaking}, neither of which is a Chainlink product. We state
their contribution as a lower bound because Etherscan caps the transfer
count it displays at $10{,}000$ per address; the true total exceeds
$380{,}000$. The contracts identified here account for a large majority but
not all of the window's single-receipt recipients. The full address list,
the funding addresses, and the Etherscan classification as retrieved on 26
August 2026 are archived with the replication package; we date the
classification because Etherscan labels are editable by their operator.

\FloatBarrier
\section{Reproducibility}
\label{app:repro}

Four details determine whether the numbers in this paper can be reproduced,
and we state them because each caused a discrepancy at some point during the
analysis.

\paragraph{Edge weights} Multiple transfers between the same ordered pair
of addresses must be summed before graph construction
(Equation~\ref{eq:edgeweight}). Standard graph libraries build a simple
directed graph in which a repeated edge overwrites its predecessor's
attribute rather than accumulating it. Because repeat interaction accounts
for $90.9\%$ and $91.7\%$ of records in the core subgraphs, an unaggregated
construction discards roughly $85\%$ of transferred volume, and the
resulting PageRank correlates with the correct one at only $0.83$ on LINK
and $0.70$ on UNI, with the top fifty addresses overlapping in $31$ and $33$
of $50$ positions.

\paragraph{Label snapshot} All supervised results are computed against a
single archived snapshot of the Etherscan tag compilation (retrieved 26
August 2026, MD5 \texttt{16792aac5afd}, $29{,}772$ entries after
case-normalisation). Querying the live source will produce a different
anchor set and therefore different AUCs.

\paragraph{Token-transfer view}
Manual verification must use Etherscan's ERC-20 token-transfer view rather
than its transaction view. A routing contract is typically called by other
contracts rather than by externally owned accounts, so its transaction count
records only the handful of calls made to it directly: among the brokers
common to both ledgers, the most structurally central shows six transactions
against several thousand token transfers. Verification that consulted the
transaction view alone would misclassify such contracts as inactive, and an
earlier pass of our own verification that inspected only the most recent
page of transfers understated the evidence available for several of them.
Etherscan also caps the displayed transfer count at $10{,}000$, which we
record as a lower bound where it binds.

\paragraph{Non-determinism} Embeddings are trained on GPU, where the
scatter operations underlying neighbourhood aggregation do not have a fixed
reduction order; seeding does not make them reproducible bit-for-bit. All
GNN results are therefore reported as means over independent
initialisations, and the broker sets as rank ensembles over ten of them. The
trained embeddings, together with a fingerprint of the feature matrix and
edge list they were derived from, are archived with the analysis code so
that the reported figures can be recomputed exactly rather than
approximately.

\section*{Declaration of Generative AI and AI-assisted Technologies in the
Writing Process}
During the preparation of this work the authors used Anthropic's Claude to
assist with drafting and editing English prose, with LaTeX formatting, and
with writing and debugging the analysis and plotting scripts described in
Sections~\ref{sec:data}--\ref{sec:exp}. All research questions, data
extraction, experimental design, model configuration, conceptual framing and
terminology, and interpretation of results were determined by the authors. All computations were executed by
the authors on their own hardware, and every reported figure was traced back
to the underlying result files. Manual on-chain verification of the broker
addresses was performed by the authors. After using this tool, the authors
reviewed and edited the content as needed and take full responsibility for
the content of this publication.

\end{document}